\documentclass[10pt,journal,compsoc]{IEEEtran}
\usepackage{graphicx}
\usepackage{pdfpages}
\usepackage{booktabs}
\usepackage{tabularx}
\usepackage{multirow}
\usepackage{amsmath,amssymb}
\usepackage{microtype}
\usepackage{url}
\usepackage{cite}
\usepackage[T1]{fontenc}
\usepackage[utf8]{inputenc}
\title{Native Dual Pools or Shared Paging? Memory Organization for Concurrent Hybrid LLM Serving}
\author{Dongwook Kang\\
Electronics and Telecommunications Research Institute (ETRI)\\
\texttt{dkang@etri.re.kr}%
\thanks{Corresponding author: Dongwook Kang (dkang@etri.re.kr). This work was supported by Institute of Information \& communications Technology Planning \& Evaluation (IITP) grant funded by the Korea government(MSIT) (No. RS-2025-022214497, Development of low-level optimization program API technology for AI semiconductors, 50\%) and by IITP grant funded by MSIT (No. RS-2023-00228255, PIM-NPU Based Processing System Software Developments  for Hyper-scale Artificial Neural Network Processing, 50\%).}}
\newif\ifTPDSsubmitted
\TPDSsubmittedtrue
\newcommand{\TPDSSubmissionNotice}{%
  \ifTPDSsubmitted
    \begin{center}
    \footnotesize\emph{This work has been submitted to the IEEE for possible publication. Copyright may be transferred without notice, after which this version may no longer be accessible.}
    \end{center}
  \fi
}

\begin{document}
\maketitle
\TPDSSubmissionNotice
\begin{abstract}
Hybrid attention--SSM serving combines growing attention KV with fixed-size recurrent and convolution state whose allocation units, reuse boundaries, and lifetimes differ. We implement four memory organizations in a modified vLLM runtime: a coarse shared-page baseline (S0), checkpoint-internal recurrent paging (S1), paging with convolution-slot packing (S2), and native dual pools (D0). Nine evaluation questions cover storage costs and serving behavior. The cumulative S0--S1--S2 refinement improves completed-output throughput and tail completion time at the main operating point. D0--S2 performance varies with memory budget, offered load, and admission headroom. Shared-prefix workloads yield larger stepwise gains; gains also persist in a no-sharing control with equal recorded input-work totals. State traces identify recurrent-checkpoint availability as the limit on joint reuse in the tested S2 runs. Extending producer-side checkpoint lifetime increases reusable depth at a fixed page budget. A fresh-seed SHARE/ISOLATE comparison estimates a positive retention interaction, with a primary confidence interval that includes zero. Retained-state continuation correctness remains unresolved. These findings motivate evaluating physical state representation together with request admission and checkpoint lifetime when designing memory management for concurrent hybrid-model serving.
\end{abstract}
\section{Introduction}\label{introduction}

Hybrid attention--SSM models make serving memory heterogeneous because attention KV, recurrent state, and convolution state differ in size, growth pattern, and reuse lifetime \cite{falcon}. A serving engine must keep these states complete and valid until later requests can reuse them. We call the amount of useful state supportable under a fixed memory budget \textbf{effective state capacity}.

PagedAttention decouples logical KV blocks from physical allocation, reducing KV fragmentation and duplication \cite{pagedattention}. A hybrid model adds fixed-size recurrent checkpoints and much smaller convolution states. Reusing a prefix requires attention state and the recurrent/convolution checkpoint to be valid at the same boundary, making the checkpoint's physical representation central to both storage efficiency and the survival of useful restart points.

Existing hybrid-serving systems study checkpoint selection, heterogeneous state allocation, and cache eviction or retention \cite{marconi}, \cite{jenga}. We investigate \textbf{how the physical representation inside a recurrent checkpoint affects concurrent serving when it is connected directly to the execution path}. Splitting a logical checkpoint across smaller physical pages and packing convolution state changes allocation waste, page mapping, ownership, and invalidation work. Our implementation and serving experiments measure the combined effect of these changes.

We implement a shared-pool refinement from coarse pages (S0) through recurrent paging (S1) to convolution-slot packing (S2), together with a native dual-pool alternative (D0).

Our first contribution is an executable hybrid-state memory organization that integrates checkpoint-internal paging and convolution-slot packing with recurrent-kernel access. S1 passes the physical member-page mapping directly to the recurrent kernel, avoiding full-checkpoint reconstruction on each decode step. S2 adds packed convolution slots with the ownership, reference, and invalidation rules that preserve the logical checkpoint as an atomic reuse unit.

Our second contribution is an evaluation of memory organization across serving conditions. We measure the S0$\rightarrow$S1$\rightarrow$S2 refinement, compare S2 with D0 across memory budgets and offered loads, and test admission headroom on independent confirmation seeds. The resulting performance curves show where each organization benefits and how admission policy changes their relative throughput.

Our third contribution is an analysis of the state histories underlying these results. Input-sharing controls relate the staged gains to reuse opportunity and recorded work. Admission-time traces identify a recurrent-checkpoint supply limit in the tested S2 runs, motivating a producer-side lifetime intervention and a fresh-seed test of its reuse-conditioned throughput effect.

Section 2 defines the hybrid-state properties and research questions. Section 3 presents the memory organizations and execution paths, Section 4 describes the experiments, and Section 5 reports Q1--Q9. Section 6 discusses design implications and limitations; Sections 7 and 8 cover related work and conclusions. Supplementary Appendices A--G provide detailed accounting, auxiliary analyses, correctness evidence, and provenance.

\section{Hybrid State Characteristics and Research Questions}\label{hybrid-state-characteristics-and-research-questions}

Attention and recurrent states impose different storage and lifetime requirements on the serving engine. This section defines the joint-reuse rule, the capacity measure used in the storage analysis, and the two research questions.

\subsection{State lifetime and joint prefix reuse}\label{state-lifetime-and-joint-prefix-reuse}

Attention KV grows with retained token history, whereas current recurrent and convolution states have fixed sizes for a given model configuration. Reuse at an earlier position requires a checkpoint saved at that position \cite{falcon}, \cite{marconi}. Logical boundaries specify these token positions; physical pages hold the corresponding state.

A prefix is reusable only when attention state and a \textbf{complete recurrent checkpoint} are valid at the same input position. In the studied model, a checkpoint includes both recurrent SSM state and convolution state. Lookup and adoption require every physical member of this joint restart point to remain valid.

\subsection{Effective state capacity}\label{effective-state-capacity}

We use \textbf{effective state capacity} for the useful state supported by a fixed memory budget. RQ1 evaluates its storage component for a fixed logical mix \(S=(x_A,x_R,x_C)\) of attention objects, recurrent checkpoints, and convolution payloads. For organization \(X\), \(M_X(S)\) is the physical allocation for this mix after applying the organization's allocation units and rounding rules. Under nominal budget \(B\), \(B/M_X(S)\) is the \textbf{allocation-capacity proxy}: a normalized storage measure for the specified mix. Actual residency also depends on pool feasibility and state lifetime.

Smaller shared pages and convolution slots reduce unused space within allocations. State-specific pools provide compact units for each type, with capacity assigned to each pool at startup. These layouts have different feasibility constraints: a static pool can fill while another pool still has free space. Section 3 derives the allocation costs and pool constraints, and the experiments track checkpoint availability, request progress, throughput, and latency during execution.

\subsection{Research questions}\label{memory-organizations-and-research-questions}

\textbf{RQ1 asks whether S0$\rightarrow$S1$\rightarrow$S2 improves allocation for the same logical state mix and whether the refinement is accompanied by better serving behavior.} The evaluation measures \(M_X(S)\), state and work histories, throughput, and latency along this cumulative implementation path.

\textbf{RQ2 asks how the complete S2 and D0 organizations compare as memory budget, offered load, and operating policy change.} Admission and lifetime experiments investigate the operating conditions and state-availability constraints associated with their measured performance.

\section{Memory Organizations, Storage Costs, and Execution}\label{memory-organizations-storage-costs-and-execution}

\begin{figure*}[!t]
\centering
\includegraphics[width=.97\textwidth,keepaspectratio]{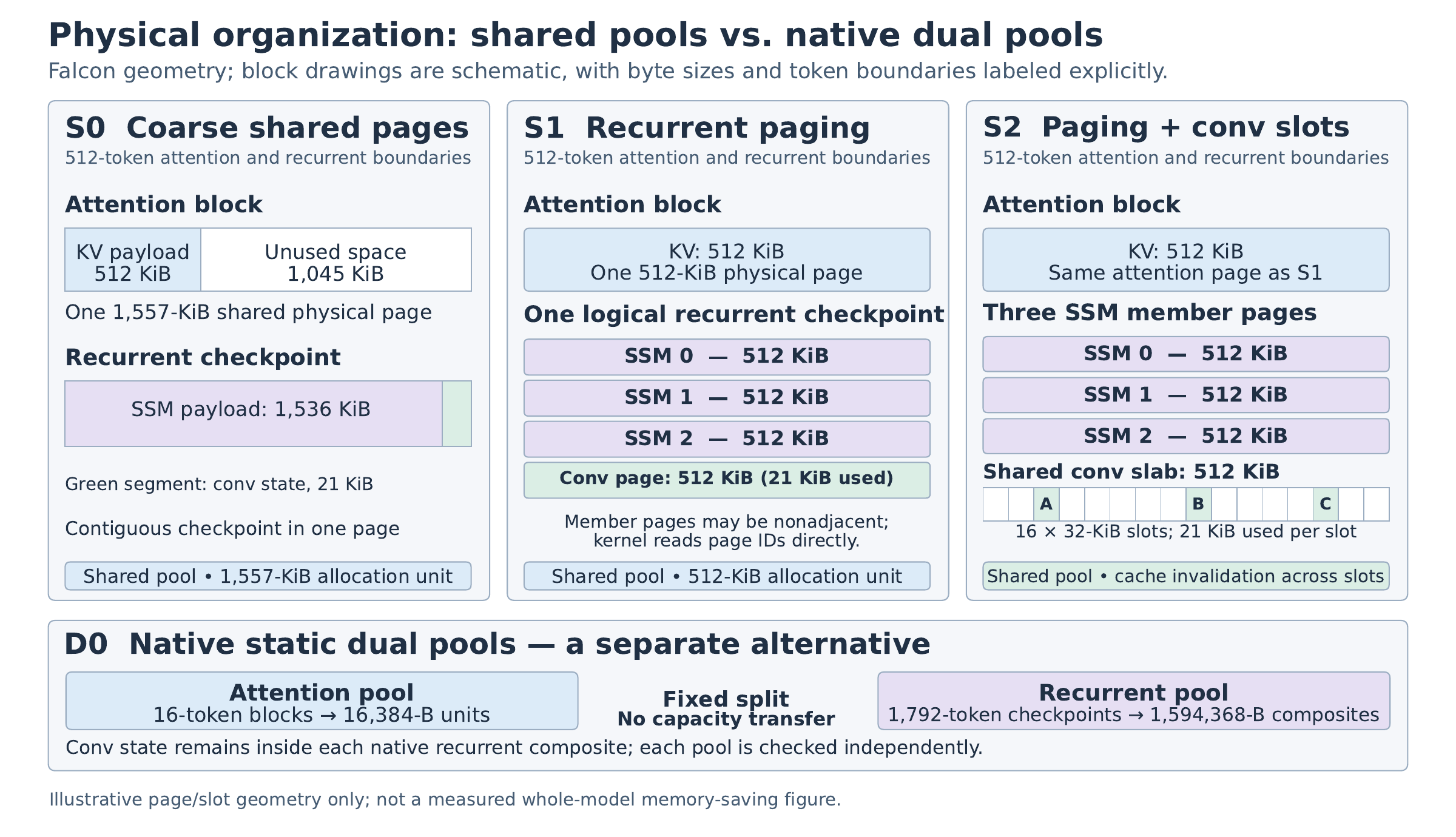}
\caption{Physical organization of the evaluated designs. S0 uses coarse shared pages; S1 introduces recurrent-checkpoint member paging with a dedicated convolution page; S2 packs convolution state into shared slabs while retaining S1 paging. D0 is a separate native dual-pool alternative with a fixed split. The illustration uses representative Falcon storage geometry, not whole-model memory savings.}
\label{fig:physical_org}
\end{figure*}

A memory organization specifies the allocation units and pools, logical-to-physical mapping, kernel access, and lifetime rules for hybrid state. Figure~\ref{fig:physical_org} illustrates the layouts; Table~\ref{tab:contract} summarizes their physical geometry: S0 uses coarse shared pages, S1 pages recurrent checkpoints, S2 adds packed convolution slots, and D0 uses static attention and recurrent pools with native block sizes. Sections~3.2--3.5 describe their execution paths and storage costs.

\begin{table*}[!t]
\caption{Memory-organization geometry. Logical boundary reports the attention-block and recurrent-checkpoint token granularity. Physical layout separates the allocation used for attention, recurrent SSM state, and recurrent convolution state. S0--S2 share the 512-token logical grid; D0 is a separate native alternative. The earlier 1,792/1,792-token uniform path explains S0's origin but is not an evaluated arm.}
\label{tab:contract}
\centering
\footnotesize
\setlength{\tabcolsep}{5.0pt}
\renewcommand{\arraystretch}{1.35}
\begin{tabularx}{\textwidth}{@{}>{\raggedright\arraybackslash}p{.095\textwidth}
>{\centering\arraybackslash}p{.080\textwidth}
>{\centering\arraybackslash}p{.090\textwidth}
>{\raggedright\arraybackslash}p{.185\textwidth}
>{\raggedright\arraybackslash}p{.225\textwidth}
>{\raggedright\arraybackslash}X@{}}
\toprule
\multirow{2}{*}{\textbf{Organization}} &
\multicolumn{2}{c}{\textbf{Logical boundary}} &
\multicolumn{3}{c}{\textbf{Physical layout}} \\
\cmidrule(lr){2-3}\cmidrule(lr){4-6}
& \textbf{Attention block} & \textbf{Recurrent checkpoint} & \textbf{Attention} & \textbf{SSM state} & \textbf{Conv state} \\
\midrule
\textbf{S0} & 512 & 512 &
one 1,557-KiB shared page\newline
for a 512-KiB KV payload &
1,536-KiB SSM payload\newline
inside one 1,557-KiB recurrent composite &
21-KiB convolution state\newline
inside the same recurrent composite \\
\textbf{S1} & 512 & 512 &
one 512-KiB shared page &
three 512-KiB member pages\newline
per logical checkpoint &
one dedicated 512-KiB page \\
\textbf{S2} & 512 & 512 &
one 512-KiB shared page &
three 512-KiB member pages\newline
per logical checkpoint &
21-KiB conv payload per 32-KiB slot;\newline
16 slots per 512-KiB slab \\
\textbf{D0} & 16 & 1,792 &
one 16-KiB native unit\newline
in the static attention pool &
1,536-KiB SSM payload\newline
inside one 1,557-KiB recurrent composite &
21-KiB convolution state\newline
inside the same recurrent composite \\
\bottomrule
\end{tabularx}
\end{table*}

\subsection{Common comparison rules}\label{common-comparison-rules}

The four organizations follow common rules for joint reuse, execution protection, and memory accounting. These rules determine when a checkpoint can be adopted, when its backing storage can be reassigned, and how the storage cost is counted.

All four organizations apply the complete-state validity rule from Section 2.1: attention, SSM, and convolution state must be valid at a common boundary. The candidate boundaries follow each organization's checkpoint grid---512 tokens for S0--S2 and the native recurrent interval for D0. Each comparison therefore uses the valid restart points supported by the evaluated organization.

An execution reference protects storage that an active request may access. After the last execution reference is released, a logical entry can remain cached; reassignment of its backing storage requires invalidating that entry. Optional retention extends protection for selected checkpoints beyond their immediate execution use. The following subsections describe these rules for coarse pages, paged composites, convolution slots, and native pools. The Supplement details preemption-time pinning and the producer-side retention policy used in the lifetime experiments.

We count each allocated physical unit once, including its unused space. A paged checkpoint contributes every required member page, and convolution states sharing one slab contribute that slab once. The formulas below cover state-memory allocation for a specified logical state set in the recorded Falcon geometry.

\subsection{S0: coarse shared pages}\label{s0-coarse-shared-pages}

S0 is the shared-pool baseline and exposes the cost of using one coarse physical unit for different state types. Both attention blocks and recurrent checkpoints use 512-token logical boundaries, but each logical object receives one 1,594,368-B (1,557-KiB) physical page. A recurrent checkpoint nearly fills that page because its representative payload is 1,536 KiB of SSM state plus 21 KiB of convolution state. A full 512-token attention block, however, contains 512 KiB of KV payload and therefore leaves 1,045 KiB unused inside the same physical unit.

The evaluated 512-token grid was inherited from the development path and held fixed throughout the shared-pool refinements. The earlier uniform-allocation path used a 1,792-token Falcon unit, set by the recurrent composite. Moving the logical boundary to 512 tokens preserves the recurrent checkpoint payload; the experiments use this grid without per-organization optimization.

S0 preserves logical checkpoint semantics while reusing this uniform pool. When execution crosses a checkpoint boundary, the runtime reads the earlier state and computes the next state into a destination page, leaving the previous checkpoint in its original page. Cached entries remain discoverable until the backing page is reassigned, at which point the corresponding logical entry must be invalidated. Thus, S0 combines flexible sharing across state types with substantial within-page waste for smaller objects. Let $x_A$ denote the number of attention-state objects and $x_R$ the number of recurrent checkpoints in a specified logical state set. Its representative allocation is
\[
\widehat B_{S0}=1557(x_A+x_R)\;\mathrm{KiB}.
\]

\subsection{S1: paging inside a recurrent checkpoint}\label{s1-paging-inside-a-recurrent-checkpoint}

S1 reduces the shared allocation unit without changing S0's 512-token logical grid. Attention uses one 512-KiB page, while one recurrent checkpoint is represented by three 512-KiB SSM member pages plus one dedicated 512-KiB convolution page. The checkpoint remains one logical object: a lookup succeeds only when all required members are valid, and reassignment of any required member invalidates the complete cached checkpoint.

The key execution change in S1 is that the recurrent kernel consumes the page mapping directly. For checkpoint boundary $b$, $\pi_b(j)$ gives the physical page of SSM member $j$, $S_{\rm page}$ is the byte stride between pages, and $\delta(j,u)$ is the in-page byte offset of element $u$ within member $j$. The abstract address relation is
\[
\operatorname{addr}_{\rm SSM}(b,j,u)=\mathrm{base}+\pi_b(j)S_{\rm page}+\delta(j,u).
\]
Here $j$ indexes physical members in the logical interface. Recurrent decode follows this mapping directly; chunked or recovery prefill may gather an initial checkpoint into a contiguous arena. Adoption protects every required member, and reassignment of any one member invalidates the complete logical checkpoint.

S1 reduces the allocation per attention block from 1,557 to 512 KiB and increases the allocation per complete recurrent checkpoint from 1,557 to 2,048 KiB. Thus,
\begin{align}
\widehat B_{S1} &= 512x_A+2048x_R\;\mathrm{KiB}, \\
\Delta \widehat B_{01} &= \widehat B_{S0}-\widehat B_{S1}
=1045x_A-491x_R\;\mathrm{KiB}.
\end{align}

S1 reduces total allocated bytes when the attention-page saving outweighs the additional checkpoint allocation for the specified state mix. The sign of \(\Delta\widehat B_{01}\) gives this condition directly.

\subsection{S2: packing convolution state into slabs}\label{s2-packing-convolution-state-into-slabs}

S2 keeps S1's paged SSM representation and removes the remaining dedicated-page cost for small convolution states. Each attention block still uses one 512-KiB page, and each checkpoint still uses three 512-KiB SSM member pages. The 21-KiB convolution state, however, occupies one 32-KiB slot in a shared 512-KiB slab containing sixteen slots. The kernel uses the slab-page identity and slot offset directly, so packing does not require reconstructing a separate contiguous convolution buffer on each step.

With slab-page index $p$ and slot index $s$, the kernel-facing line identifier is $16p+s$ and the line stride is 32 KiB. An explicit invalid sentinel represents absent convolution state and prevents missing entries from aliasing page zero or a negative-offset store.

Sharing a slab couples the lifetimes of its cached objects. Active references protect the relevant slot and its checkpoint relationship, while the page can contain both active and cache-only slots. Before a slab page is reassigned, every cached checkpoint using one of its slots must be invalidated. Supplementary Appendix D.2--D.3 specifies the reference-count and invalidation rules.

S2's storage cost depends on both checkpoint count and slab occupancy. Let $m$ denote the number of allocated convolution slabs backing the $x_R$ checkpoints. Then
\begin{align}
\widehat B_{S2} &= 512x_A+1536x_R+512m\;\mathrm{KiB},
\quad m\ge \lceil x_R/16\rceil, \\
\Delta \widehat B_{12} &= \widehat B_{S1}-\widehat B_{S2}
=512(x_R-m)\;\mathrm{KiB}.
\end{align}
The last term gives the S1$\rightarrow$S2 saving for the specified object set. Allocation history can leave partially occupied slabs, so $m$ can exceed its lower bound.

\subsection{D0: native units in two static pools}\label{d0-native-units-in-two-static-pools}

D0 is a separate complete organization that avoids both S0's coarse attention padding and S1's recurrent page rounding. It allocates attention in native 16-token, 16,384-B units and recurrent checkpoints as native 1,594,368-B contiguous composites at 1,792-token intervals. These objects draw from separate attention and recurrent pools whose capacities are fixed at startup; unused capacity in one pool is not transferred to satisfy demand in the other.

D0's compact state-specific units are subject to two pool-feasibility constraints. Let \(x^{D0}_{A,16}\) and \(x^{D0}_{R}\) count native attention blocks and recurrent checkpoints. For pool budgets \(M_A\) and \(M_R\), a state mix is feasible only when both \[
16x^{D0}_{A,16}\le M_A
\quad\text{and}\quad
1557x^{D0}_{R}\le M_R
\] hold in KiB units. Both constraints must hold even when the combined state allocation fits within \(M_A+M_R\). D0's native checkpoint grid and allocation checks are part of the configuration evaluated against S2.

\subsection{Comparison contract}\label{comparison-contract}

S0--S2 use the same 512-token attention-block and recurrent-checkpoint grid; physical representation, kernel access, and lifecycle support change cumulatively along this path. D0 uses native 16-token attention blocks and 1,792-token recurrent checkpoints with a fixed split between attention and recurrent pools. The S2--D0 comparison thus includes differences in checkpoint grids, allocation units, and pool-feasibility checks.

\section{Experimental Design and Measurement}\label{experimental-design-and-measurement}

Table~\ref{tab:qmap} maps Q1--Q9 to the RQ1 shared-pool refinement, the RQ2 S2--D0 comparison, and the state-availability and lifetime experiments. Each campaign has its own sample and analysis. Supplementary Appendices B, C, E, and G record the runs and statistical procedures.

\begin{table*}[!t]
\caption{Q1--Q9 experiment map. $\lambda$ is offered request rate, $q$ the common-prefix fraction, and cap the configured sequence-count limit. Budget labels are planner-matched nominal budgets, not GiB values. Q7--Q9 are separate evidence stages and their samples are not pooled. For Q9, the preregistered primary analysis uses the ledger timebase; request-span results are a sensitivity analysis.}
\label{tab:qmap}
\centering
\footnotesize
\setlength{\tabcolsep}{3.0pt}
\renewcommand{\arraystretch}{1.22}
\begin{tabularx}{\textwidth}{@{}>{\raggedright\arraybackslash}p{.030\textwidth}>{\raggedright\arraybackslash}p{.185\textwidth}>{\raggedright\arraybackslash}p{.105\textwidth}>{\raggedright\arraybackslash}p{.250\textwidth}>{\raggedright\arraybackslash}p{.135\textwidth}>{\raggedright\arraybackslash}X@{}}
\toprule
\textbf{Q} & \textbf{Question / purpose} & \textbf{Compared arms} & \textbf{Fixed anchor} & \textbf{Varied factor} & \textbf{Reported readouts} \\ \midrule
\multicolumn{6}{@{}l}{\textbf{RQ1 --- representation refinement and input sharing}} \\
Q1 & Main-load stepwise comparison (Exp. 1a): does the cumulative shared refinement improve end-to-end performance? & S0\newline S1\newline S2 &
$\lambda=3$ req/s\newline
cap 63\newline
$q=1$\newline
seeds 42000--42011\newline
512-token boundaries &
memory organization\newline
b63 / b83 / b108 &
throughput\newline
paired step change\newline
E2E p99 \\
Q1 & Low-load throughput and latency check (Exp. 1b) & S0\newline S1\newline S2 &
$\lambda=.8$ req/s\newline
seeds 0--2\newline
512-token boundaries &
memory organization\newline
b63 / cap 63\newline
b108 / cap 109 &
throughput\newline
TTFT / E2E p50,p99\newline
cache use / waiting\newline
preemption \\
Q2 & What runtime work/state changes accompany the Q1 path? & S0\newline S1\newline S2 &
b63 / cap 63, $\lambda=3$\newline
b108 / cap 109, $\lambda=4.5$\newline
seeds 0--2 &
memory organization &
recorded input work\newline
running / waiting\newline
cache use\newline
preemption\newline
throughput \\
Q2 & Is lower recorded input work necessary for the stepwise gain? (Exp. 2c counterexample) & S0\newline S1\newline S2 &
b83\newline
$\lambda=3$ req/s\newline
cap 63\newline
$q=0$\newline
seeds 42000--42011 &
memory organization &
initial-input component\newline
recorded token total\newline
preemption\newline
throughput \\
Q3 & Does common-prefix sharing interact with memory organization? & S0\newline S1\newline S2\newline D0 &
b83\newline
$\lambda=3$ req/s\newline
cap 63\newline
seeds 42000--42011\newline
cache reuse enabled &
$q=0$\newline
$q=.5$\newline
$q=1$ &
relative throughput\newline
organization $\times$ sharing interaction \\
\addlinespace[2pt]
\midrule
\multicolumn{6}{@{}l}{\textbf{RQ2 --- complete-organization operating conditions}} \\
Q4 & How do all four organizations compare across memory budget? & S0 / S1 / S2\newline (reused Q1 runs)\newline D0 &
$\lambda=3$ req/s\newline
cap 63\newline
$q=1$ &
b63\newline b83\newline b108 &
throughput by budget\newline
direct paired D0/S2 contrast \\
Q5 & How does the ranking change with offered load? & D0\newline S2 &
$q=1$\newline
b63 / cap 63\newline
b108 / cap 109\newline
Poisson arrivals &
$\lambda=.8$--3.0\newline
$\lambda=.8$--4.5 &
throughput vs. offered load\newline
D0/S2 gap\newline
expanded high-load tests kept separate \\
Q6 & Is the D0--S2 result sensitive to admission headroom? & D0\newline S2 &
b63 / b83 / b108\newline
$\lambda=3$ req/s\newline
cap 63\newline
$q=1$\newline
fresh seeds 76400--76411 &
$f=0$\newline
$.10$\newline
$.19$\newline
$.27$\newline
$.35$ &
throughput\newline
D0/S2 gap\newline
TTFT / E2E p99\newline
effect of preselected headroom vs. $f=0$ \\
\addlinespace[2pt]
\midrule
\multicolumn{6}{@{}l}{\textbf{Mechanism and confirmation --- separate campaigns; not pooled}} \\
Q7 & Which state limits joint reuse depth? & D0\newline S2 &
b83\newline
$\lambda=3$ req/s\newline
cap 63\newline
300 requests\newline
seeds 76600--76602\newline
matched trace OFF / ON &
$f=0$\newline
$f=.19$ &
attention availability $A_{\max}$\newline
$R_{\max}$\newline
$J_{\rm enter}$\newline
$J_{\rm adopt}$\newline
trace ON/OFF throughput check\newline
(90\% CI within $\pm3$\%) \\
Q8 & Does longer checkpoint lifetime restore reusable depth? & S2\newline S2+Retention &
254 pages\newline
$\lambda=3$ req/s\newline
cap 63\newline
$f=0$\newline
SHARE\newline
seeds 79010--79012 &
producer-checkpoint lifetime &
$J_{\rm adopt}$ / $R_{\max}$\newline
absolute depth / paired change\newline
$A_{\max}$: offline diagnostic\newline
throughput \\
Q9 & Is the retention gain conditioned on cross-request reuse? & S2 / S2+Retention\newline SHARE / ISOLATE &
254 pages\newline
$\lambda=3$ req/s\newline
cap 63\newline
$f=0$\newline
fresh seeds 88001--88006\newline
balanced order &
lifetime $\times$ reuse condition &
absolute ledger throughput (descriptive)\newline
primary interaction $K_{\rm ret}$\newline
with 95\% CI\newline
secondary $\Delta_{\rm SHARE}$ / $\Delta_{\rm ISOLATE}$ \\
\bottomrule
\end{tabularx}
\end{table*}

\subsection{Experimental setting and performance measures}\label{common-environment-and-measurement-contract}

All main comparisons run Falcon-H1-7B-Instruct with BF16 and tensor parallelism 1 on a single NVIDIA A100-PCIE-40GB in the modified vLLM V1 runtime. We vary state-memory capacity through the planner targets b63, b83, and b108. Each organization converts the same nominal byte target into an integer count of its own blocks or pages, with D0's attention/recurrent split fixed before the comparison. These \textbf{planner-matched memory budgets} retain the allocation rounding of each design. The available campaign-specific software and workload records are provided in Appendices B and F.

We use workload seeds as independent repeats, pairing configurations by seed and common generator settings. Throughput counts successfully completed output tokens over the measurement interval, usually the span from the first request start to the last successful response. In the Q1--Q6 plots, absolute throughput is averaged arithmetically across runs; relative effects and S0-normalized ratios are formed within seeds before averaging. A reported mean p99 for TTFT or E2E averages the run-level percentiles over those seeds.

Q9 uses the ledger interval for its preregistered primary throughput analysis. This interval includes benchmark orchestration and completion polling, with server boot and shutdown outside the interval. Request-span throughput provides the timing sensitivity analysis (Appendix G.3--G.4).

\subsection{Shared-path refinement and input sharing: Q1-Q3}\label{shared-path-refinement-and-input-sharing-q1-q3}

Q1--Q3 analyze the cumulative S0$\rightarrow$S1$\rightarrow$S2 path under ShareGPT-V3-derived workloads (eight shared contexts, approximately 3,332 input tokens, model limit 8,192, generation cap 1,024). Table~\ref{tab:qmap} specifies the seeds, budgets, arrivals, and caps. Q1 measures throughput and latency, Q2 records work/progress and tests the $q=0$ counterexample, and Q3 varies sharing. Shared runs are not additional independent repetitions.

Q2 measures recorded initial-input tokens with $C_{\rm once}$ and periodically samples the numbers of running and waiting requests. These serving-level counters do not measure total GPU work (Appendix B.4).

In the Exp. 2c no-sharing control, $q=0$ removes the common-prefix reuse opportunity while cache creation and lookup remain enabled. This control measures the refinement's performance when recorded input-work totals are equal. Figure~\ref{fig:q2work}(b) shows a matched-seed normalized index. The displayed $q=0$ indices are reconstructed from archived seedwise contrasts; the S2/S0 ratio is formed within each seed before averaging over the same twelve seeds (Appendix G.1).

\subsection{Budget and offered-load comparisons: Q4-Q5}\label{budget-and-offered-load-comparisons-q4-q5}

Q4 reuses the Q1 S0/S1/S2 runs and adds D0 at fixed 3 req/s and cap 63, with D0/S2 as the direct inferential contrast. Q5 varies Poisson offered load for b63/cap 63 and b108/cap 109, using common seeds 0--2 for the curves. Expanded high-load comparisons retain their own sample sizes. The Q5 curves characterize finite request streams at these two budget--cap configurations.

\subsection{Admission headroom: Q6}\label{admission-headroom-q6}

Q6 varies the \textbf{admission headroom fraction} \texttt{f} within each physical organization. For total block count \texttt{B}, the scheduler requires an additional headroom of \texttt{H=floor(fB)} blocks when admitting a waiting or preempted request; this margin is a feasibility condition applied before allocation. Its byte value depends on each organization's allocation unit, and the resulting execution determines the running set.

Q6 sweeps the five headroom fractions and 12 fresh seeds specified in Table~\ref{tab:qmap}. Separate development runs selected one policy value for each organization/budget pair; the confirmation compares that value with $f=0$ (Appendix B.12).

\subsection{Checkpoint lifetime and reuse: Q7--Q9}\label{state-availability-checkpoint-lifetime-and-reuse-q7-q9}

Q7 traces joint reuse at admission in b83: the attention prefix must reach a valid recurrent checkpoint. We reconstruct queue-entry $J_{\rm enter}$ offline and record $J_{\rm adopt}$ in the scheduler; lookup-time attention depth $A_{\max}$ and recurrent depth $R_{\max}$ follow the M1 supply-mask accounting bound (Appendix G.2). Trace-ON histories are paired with trace-OFF throughput runs. For each organization--headroom pair, a two-sided 90\% CI over three paired seedwise log-throughput effects must lie within $\pm3\%$ before offline reanalysis (Appendices C.1 and D.4).

Q8 changes checkpoint lifetime within S2's fixed 254-page budget. S2+Retention exempts 512-token checkpoints meeting the $K=2$ distinct-request demand threshold from release while the producer runs; ordinary caching and pinning apply after completion or preemption. Under SHARE, the depth comparison averages 300 matched INITIAL requests per arm--seed, excluding RESUME episodes (Fig.~\ref{fig:retention}(b); Appendix G.2).

Q9 crosses S2/S2+Retention with SHARE/ISOLATE in balanced order over fresh seeds 88001--88006. SHARE omits \texttt{cache\_salt}; ISOLATE uses a stable request-unique salt to block cross-request reuse while retaining caching and same-request resume. All four conditions use the same model, runtime, and workload generator at the fixed settings in Table~\ref{tab:qmap}. Recorded prompt lengths, contexts, and output targets match within seeds; full token IDs and planned-arrival vectors were not logged.

For seed $s$, policy $p$, and reuse mode $m$, let $G_{s,p,m}$ be ledger throughput and $d_m(s)=\ln(G_{s,\mathrm{Ret},m}/G_{s,\mathrm{S2},m})$. With $\langle\cdot\rangle_s$ denoting the six-seed mean, the displayed arm value and reported effects are
\begin{align*}
\widetilde G_{p,m}&=\exp\!\left(\langle\ln G_{s,p,m}\rangle_s\right),\\
\Delta_m(\%)&=100\left[\exp\!\left(\langle d_m(s)\rangle_s\right)-1\right],\\
K_{\rm ret}(\%)&=100\left[\exp\!\left(\langle d_{\rm SHARE}(s)-d_{\rm ISOLATE}(s)\rangle_s\right)-1\right].
\end{align*}
Ledger $K_{\rm ret}$ is primary, with a paired-$t$ 95\% CI on the six seedwise log-effect differences ($df=5$). Bars are descriptive geometric means; within-mode effects are secondary, with no ISOLATE equivalence test, and request-span results assess timing sensitivity (Appendix G.4).

The observe-only logs record 48 distinct $K=2$-eligible hashes and 283--284 filter-kept requests per S2+Retention--SHARE run, compared with zero eligible hashes under ISOLATE. Six resume re-entries of four requests occurred under SHARE, each with a 3,072-token joint-prefix hit. Physical free-skip counts and adopted-state ownership were not recorded, and ISOLATE self-resume was not exercised.

\section{Performance Results and Reusable-State Analysis}\label{performance-results-and-reusable-state-analysis}

We report Q1--Q9 in the order of Table~\ref{tab:qmap}. Figures~\ref{fig:step}--\ref{fig:retention} summarize the results; the separate Q4 fixed-load budget comparison is discussed in Section~\ref{memory-budget-and-offered-load-q4q5}, whereas Fig.~\ref{fig:load} depicts Q5. Source tables and sensitivity analyses are in Appendices E and G.

\subsection{Stepwise performance changes: Q1}\label{stepwise-performance-changes-q1}

\begin{figure*}[!t]
\centering
\includegraphics[width=.98\textwidth,height=3.05in,keepaspectratio,trim=0pt 15pt 0pt 0pt,clip]{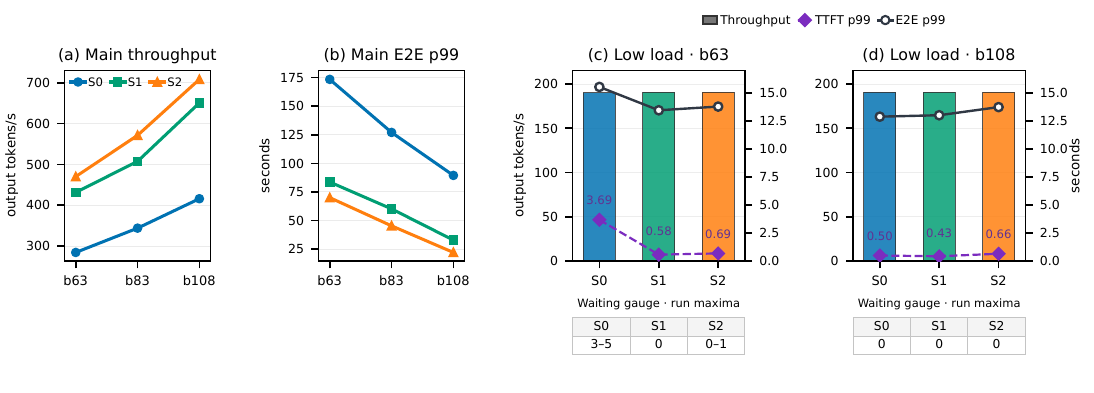}
\vspace{-11pt}
\caption{Q1 performance at main and low offered load. Panels (a,b) report Exp. 1a at 3 req/s over twelve paired seeds per cell; (c,d) show Exp. 1b at 0.8 req/s over seeds 0--2 for b63/cap 63 and b108/cap 109. In (c,d), organization-colored bars use the left throughput axis; diamonds and circles use the right latency axis for TTFT p99 and E2E p99. Throughput values are arithmetic run means, whereas reported relative gains average within-seed ratios. Latency values average run-level p99s. The micro-tables give ranges of sampled run-level maximum waiting-request counts.}
\label{fig:step}
\end{figure*}

The cumulative S0$\rightarrow$S1$\rightarrow$S2 refinement improves both throughput and tail completion time at the main 3 req/s operating point. Figure~\ref{fig:step}(a,b) shows the twelve-seed Exp. 1a results at b63, b83, and b108, while Fig.~\ref{fig:step}(c,d) shows the separate three-seed low-load Exp. 1b results. Mean seedwise S1/S0 throughput gains are 51.784\%, 47.568\%, and 56.235\%, and the additional S2/S1 gains are 8.929\%, 12.628\%, and 9.972\%, respectively. All six stepwise contrasts remain distinguishable after the original 22-test Holm correction. At b108, S2/S1 is positive in ten seeds and negative in two. The b108 S1 results include a rerun; Appendices B.12 and E.14 give the seed-exclusion sensitivity.

At 0.8 req/s, S0, S1, and S2 all deliver about 190 completed-output tokens/s at both budgets, with larger differences in tail latency at b63 (Fig.~\ref{fig:step}(c,d)). At b63, S0 has a mean run-level TTFT p99 of 3.687 s, compared with 0.578 s for S1 and 0.686 s for S2. Its sampled maximum waiting-request count ranges from 3--5, compared with zero for S1 and 0--1 for S2. At b108, all three organizations have zero sampled waiting and mean run-level TTFT p99 values below one second (0.500, 0.430, and 0.657 s). Waiting counts come from a periodic aggregate gauge with no request identifiers.

\subsection{Input work, request progress, and sharing: Q2--Q3}\label{input-work-request-progress-and-sharing-q2q3}

\begin{figure}[!t]
\centering
\includegraphics[width=.99\columnwidth,keepaspectratio]{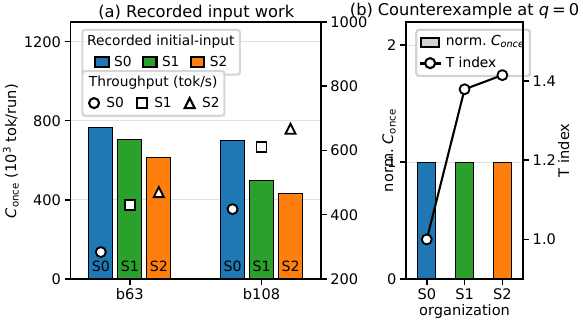}
\vspace{-6pt}
\caption{Q2 recorded work and the no-sharing control. (a) Bars: $C_{\rm once}$ ($10^3$ tokens/run, left); hollow symbols: throughput (tokens/s, right). Each group is a three-run mean: b63/cap 63 at 3 req/s and b108/cap 109 at 4.5 req/s. (b) At b83, $q=0$, 3 req/s, and cap 63, bars show normalized $C_{\rm once}=1$; the line shows mean within-seed S0-normalized throughput over twelve paired seeds. Both normalized quantities in (b) are dimensionless. Appendix G.1 provides exact values and reconstruction.}
\label{fig:q2work}
\end{figure}

At the two traced operating points, the stepwise throughput gain accompanies lower recorded initial-input work and less waiting (Fig.~\ref{fig:q2work}(a)). At b63 and 3 req/s, the recorded initial-input component falls from about 766k tokens/run in S0 to 612k in S2, while median waiting falls from 78.8 to 25.7 and median running requests rise from 6.7 to 17.0. At b108 and 4.5 req/s, the corresponding initial-input values fall from about 698k to 430k and median waiting from 67.7 to 8.5.

Throughput also improves in the no-shared-context Q2 control, where recorded work and preemptions remain constant. At b83, $q=0$, and cap 63, S0/S1/S2 have identical recorded initial-input and total-token counts and zero preemptions. Their arithmetic mean throughputs are 293.836/405.376/415.862 output tokens/s (supplementary Table G-1). Figure~\ref{fig:q2work}(b) presents the corresponding mean within-seed S0-normalized indices, defined in Appendix G.1.

\begin{figure}[!t]
\centering
\includegraphics[width=.98\columnwidth,keepaspectratio,trim=0pt 9pt 0pt 8pt,clip]{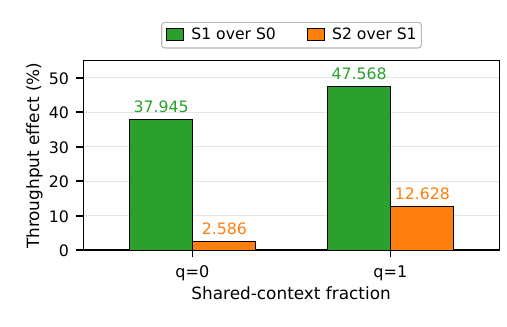}
\vspace{-10pt}
\caption{Q3 shared-context endpoints at b83 (3 req/s, cap 63; 12 matched seeds). Bars show mean seedwise throughput effects at $q=0,1$. The endpoint interactions are +9.623 percentage points for S1/S0 and +10.043 for S2/S1 (both Holm-adjusted $p=0.01074$). The $q=0$ S2/S1 value is 2.586\%. Appendix G, Table G-2 contains $q=0.5$ and D0/S2 results.}
\label{fig:q3sharing}
\end{figure}

Q3 shows larger endpoint effects with fully shared context (Fig.~\ref{fig:q3sharing}). Increasing that fraction from 0 to 1 strengthens the reported mean S1/S0 effect by 9.623 percentage points and S2/S1 by 10.043 points; both endpoint interactions have Holm-adjusted $p=0.01074$. Both refinement steps therefore yield larger gains in the fully shared workload (supplementary Table G-2).

\subsection{Memory budget and offered load: Q4--Q5}\label{memory-budget-and-offered-load-q4q5}

The D0--S2 ordering changes across the tested memory budgets. At the common 3 req/s, cap-63 comparison with zero headroom, the mean paired D0/S2 difference is +4.978\% at b63, -1.728\% at b83, and -12.556\% at b108; the b63 and b108 contrasts have Holm-adjusted \(p=0.01074\), while the b83 contrast is inconclusive. Q4 holds cap 63 fixed at every budget; the Q5 b108 load sweep below uses cap 109. Budget and auxiliary load details are in supplementary Appendix E.2.

\begin{figure*}[!t]
\centering
\includegraphics[width=.87\textwidth,keepaspectratio]{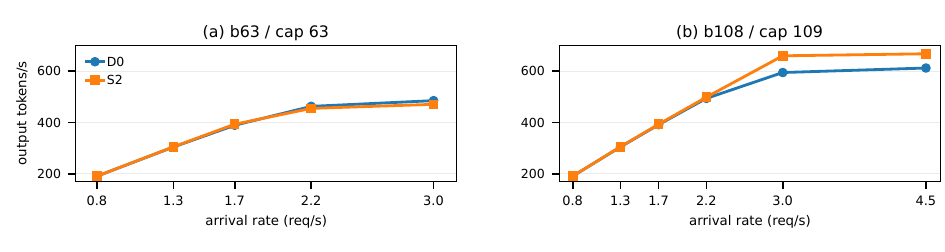}
\vspace{-4pt}
\caption{Q5 D0--S2 offered-load sweeps at (a) b63/cap 63 and (b) b108/cap 109, with $q=1$. Curves show arithmetic means of run-level throughput over common seeds 0--2. Expanded six- or nine-seed high-load tests are separate comparisons. The Q4 budget sweep uses fixed cap 63.}
\label{fig:load}
\end{figure*}

Offered load further changes the D0--S2 gap, and the direction depends on the budget configuration. Figure~\ref{fig:load} uses the common seed 0--2 means for the full curves. At b63/cap 63 the two organizations are nearly identical at low rates and D0 becomes slightly faster at the upper end of the curve. At b108/cap 109, S2 becomes faster at 3.0 and 4.5 req/s. At b63, the expanded 3.0 req/s comparison gives a mean paired D0/S2 difference of +3.64\% over six seeds (Holm \(p=0.3750\)), while b108 at 4.5 req/s gives -5.97\% over nine seeds (Holm \(p=0.0391\)). Among the common six b108 seeds, the gap changes from -13.17\% at 3.0 req/s to -4.74\% at 4.5 req/s, giving S2 a smaller relative advantage at the higher load.

These curves summarize throughput over finite request streams at the stated memory budgets and sequence-count caps (Appendices E.2--E.5). This dependence on memory budget and offered load motivates examining admission-headroom sensitivity without changing the underlying physical organizations.

\subsection{Admission-headroom sensitivity: Q6}\label{admission-headroom-sensitivity-q6}

\begin{figure*}[!t]
\centering
\includegraphics[width=.92\textwidth,trim=0pt 8pt 0pt 9pt,clip]{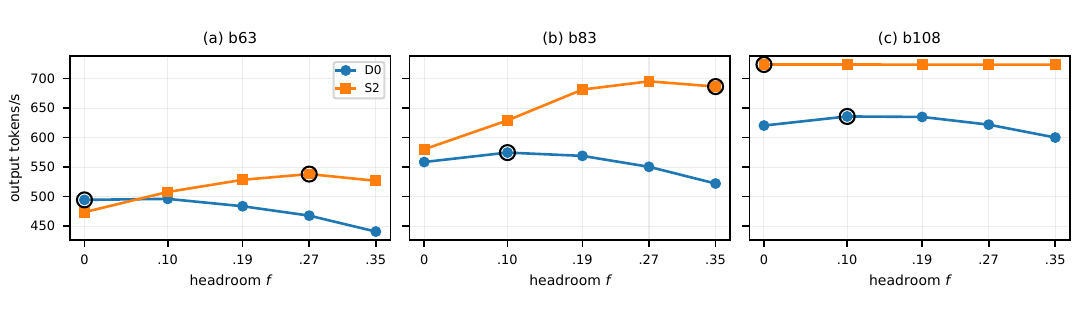}
\vspace{-4pt}
\caption{Q6 admission-headroom sensitivity at 3 req/s, cap 63, and $q=1$. Curves are arithmetic means of run-level throughput over fresh seeds 76400--76411. Open rings mark organization--budget-specific headroom values chosen before independent confirmation (Appendix B.12).}
\label{fig:headroom}
\end{figure*}

The throughput response to admission headroom varies by organization and budget. Figure~\ref{fig:headroom} reports the independent confirmation over 12 fresh seeds per cell for \(f\in\{0,0.10,0.19,0.27,0.35\}\). S2 benefits from positive headroom at b63 and b83, whereas its b108 curve is nearly flat; D0 peaks at small positive values for b83/b108 and declines as headroom increases further.

Several preselected settings reproduce positive effects on the fresh seeds, and the policy can change the D0--S2 ranking. Relative to its own \(f=0\), preselected S2 gains 13.61\% at b63 (\(f=0.27\)) and 18.81\% at b83 (\(f=0.35\)); D0 gains 2.86\% and 2.50\% at b83 and b108 with \(f=0.10\). With common \(f=0\), D0 is higher at b63 and S2 at b108, while b83 is inconclusive. With common \(f=0.19\), or with each organization's preselected value, S2 is higher at all three budgets. In S2-b83, the displayed mean at \(f=0.27\) exceeds that at the preselected \(f=0.35\). The preselected $f=0.35$ remains the confirmation setting; the available data do not resolve whether $f=0.27$ improves throughput by more than 3\% (Appendix B.12).

\subsection{Checkpoint lifetime and reuse: Q7--Q9}\label{state-availability-checkpoint-lifetime-and-reuse-q7q9}

\begin{figure*}[!t]
\centering
\includegraphics[width=.98\textwidth,height=2.75in,keepaspectratio]{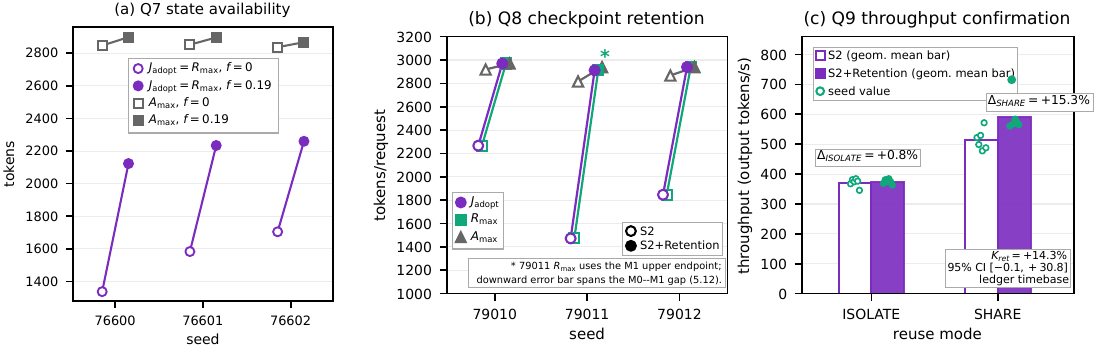}
\vspace{-4pt}
\caption{Q7--Q9 checkpoint availability and retention results. Panels (a,b) show mean absolute request depths. (a) Open/filled markers denote $f=0/0.19$ in S2; purple circles combine $J_{\rm adopt}=R_{\max}$, and gray squares show $A_{\max}$. (b) Open/filled markers denote S2/S2+Retention at 254 pages; each value averages 300 matched INITIAL requests, excluding RESUME episodes. $R_{\max}$ and $A_{\max}$ use lookup-time M1; the seed-79011 lower $R_{\max}$ bar spans its 5.12-token M0--M1 measurement bracket. Horizontal offsets only separate overlapping metrics. Full brackets are in Appendix G.2. (c) Bars are six-seed geometric-mean ledger throughputs; dots show the 24 run-level observations (Table G-5). Group effects are secondary; the boxed primary $K_{\rm ret}$ is +14.3\% (95\% CI $[-0.1,+30.8]$), which includes zero.}
\label{fig:retention}
\end{figure*}

Q7--Q9 examine recurrent-checkpoint availability and the effect of extending producer-side checkpoint lifetime. Figure~\ref{fig:retention}(a) shows Q7 seedwise state availability under two headroom settings; panel (b) compares absolute Q8 state depths in S2 and S2+Retention; panel (c) compares absolute Q9 ledger throughput under shared and isolated cross-request reuse.

In Q7, the available recurrent checkpoints set the adopted joint reuse depth in all six tested S2 cells. Let \(J\) be the deepest complete checkpoint usable within the available attention prefix and \(R_{\max}\) the deepest available recurrent checkpoint independently of that attention window. Their difference at adoption is \(Gap_R=R_{\max}-J_{\rm adopt}\). For \(f=0\) and \(f=0.19\), \(J_{\rm adopt}=R_{\max}\) and \(Gap_R=0\) in all three seeds. The paired increases are 785.1, 650.2, and 554.7 tokens when headroom rises from 0 to 0.19 (Appendix G.2).

At the same 254-page budget, producer-side checkpoint retention increases recurrent availability and joint reuse in Q8. This experiment compares S2 with S2+Retention under cross-request reuse. Figure~\ref{fig:retention}(b) shows the per-arm INITIAL-request means; supplementary Tables G-3--G-4 give their M0--M1 brackets and paired changes. Across seeds 79010--79012, \(\Delta J_{\rm adopt}\) is approximately +703, +1439, and +1092 tokens/request, while \(\Delta R_{\max}\) is approximately +703, +1444, and +1092. Attention availability also increases, by approximately +51, +123, and +72 tokens/request in \(A_{\max}\), respectively, substantially less than the recurrent/joint-depth gains. The single ambiguity in seed 79011 bounds \(\Delta Gap_R\) within {[}0,+5.12{]} tokens/request. All three seedwise throughput effects are positive; the aggregate effects are +17.8\% on the ledger timebase and +23.5\% on the request-span timebase. The reported n=3 performance interval includes zero.

The fresh-seed Q9 confirmation estimates a larger retention benefit under shared reuse. Figure~\ref{fig:retention}(c) compares six-seed geometric-mean ledger throughput: S2 versus S2+Retention yields 371.42 versus 374.52 output tokens/s under ISOLATE, and 513.70 versus 592.07 under SHARE. Dots show run-level throughput. Under SHARE, the secondary ledger effect is +15.3\% with 95\% CI {[}+0.8,+31.8{]}; under ISOLATE it is +0.8\% {[}-1.3,+3.0{]}. The primary ledger interaction \(K_{\rm ret}\), computed from the seedwise log-effect difference before converting back to percent, is +14.3\% (95\% CI {[}-0.1,+30.8{]}) and is positive in all six seeds. The 95\% CI includes zero; the primary interaction is therefore statistically inconclusive.

\section{Discussion and Limitations}\label{discussion-interpretation-boundaries-and-scope}

Several checkpoints can share an S2 convolution slab, which couples their reclamation. Active references protect the members needed by running requests; before reassigning a slab, the runtime invalidates every dependent cached checkpoint. The compact representation therefore requires tracking dependencies across page and slot boundaries, alongside the kernels' direct access to those allocations.

Active requests and reusable checkpoints compete for capacity within the selected pool organization. The headroom-sensitive D0--S2 ranking ties layout selection to the operating policy. Keeping an earlier recurrent checkpoint can preserve a restart point for an available attention prefix, while occupying memory for longer. Salt-based isolation can change state supply and memory pressure as well as cross-request hits, so the SHARE/ISOLATE response reflects the working set created by that intervention.

The results cover Falcon-H1-7B-Instruct on A100 40GB at the evaluated budgets, loads, and sharing patterns. Applying the layout to another state geometry requires assessing its allocation and execution costs. Appendices B.12, E, and F give the available run identities and analysis artifacts; some historical runs lack exact commits, run-time usable-memory fields, or complete raw records.

Clean continuation comparisons found output differences between selected retained-state paths and independent recomputation; full-workload isolation also produced variability with cross-request retained-state reuse disabled. The cause remains unresolved, leaving exact-continuation correctness open. Appendix D.5 records the reference-repeatability and treatment-fidelity conditions needed to interpret those comparisons.

\section{Related Work}\label{related-work-and-positioning}

\subsection{Physical paging and prefix reuse}\label{physical-paging-and-prefix-reuse}

PagedAttention separates logical KV blocks from physical allocation \cite{pagedattention}; vAttention maintains contiguous virtual addresses over dynamically managed physical KV memory \cite{vattention}; and SGLang/RadixAttention combines prefix-tree caching with cache-aware ordering \cite{radix}. S1 brings physical paging inside a recurrent checkpoint and passes its member-page mapping directly to the recurrent kernel. S2 extends the representation with convolution-state slots in shared slabs.

\subsection{Hybrid checkpoint management and retention}\label{hybrid-checkpoint-management-and-retention}

Marconi manages cache admission and eviction for hybrid checkpoints \cite{marconi}. Sparse Prefix Caching selects logical checkpoint positions under overlap and budget constraints \cite{sparseprefix}, while Jenga combines heterogeneous allocation, layer-specific caching and eviction, and request dependencies \cite{jenga}. AVMP and SGLang's hybrid-memory implementation provide further approaches to flexible state pools \cite{avmp}, \cite{sglanghybrid}. Our design operates within a complete recurrent checkpoint, exposing its SSM member pages and convolution slot to the kernel.

SegLen considers recency together with replay distance to a valid recurrent ancestor \cite{seglen}, and vLLM's hybrid-serving implementation describes interval and selective checkpoint retention \cite{vllmhybrid}. The producer-side retention setting extends checkpoint lifetime while keeping S2's page and slot layout fixed.

\subsection{Request admission and scheduling}\label{request-admission-and-scheduling}

Orca provides iteration-level scheduling \cite{orca}, and Sarathi-Serve uses chunked prefill and batch construction to reduce prefill/decode interference \cite{sarathi}. Marconi's cache-admission policy selects checkpoint state for retention \cite{marconi}. The headroom policy studied here acts on waiting or preempted requests, requiring a free-memory margin before they join the running set.

\section{Conclusions and Future Work}\label{conclusions-and-future-work}

This paper integrates checkpoint-internal paging and convolution-state packing into a hybrid-model serving engine. Along the S0$\rightarrow$S1$\rightarrow$S2 path, S1 reduces coarse attention allocation and S2 lets small convolution states share pages. These refinements improve throughput and tail completion time at the main-load operating points. Common-prefix sharing strengthens the gains, and the no-sharing control retains gains with unchanged recorded input-work counters. The S2--D0 ordering varies with memory budget, offered load, and admission headroom. At a fixed page budget, producer-side lifetime extension increases recurrent availability and adopted joint depth.

Future work should measure paging and packing overhead at matched usable capacity, evaluate joint admission--retention policies on additional models and devices, and investigate continuation divergence using controlled recomputation and repeatability tests.

\section*{Acknowledgment}
OpenAI ChatGPT was used to assist with drafting and revising manuscript text in Sections 1--8 based on the author's technical descriptions and experimental results, preparing LaTeX source files, creating and revising the memory-organization diagram in Fig. 1, and developing and revising plotting scripts for the experimental figures (Figs. 2--7) and their captions. The author reviewed the AI-assisted material and is responsible for the accuracy of the reported data, analyses, figures, and technical conclusions.

\clearpage
\pagestyle{empty}


\section*{Supplementary material}

# Appendix A. Conditional Storage Capacity and Addressing

## A.1 Conditional Capacity of the Native Two-Pool Design

A fixed state mix makes D0's two pool limits explicit without turning the calculation into a request-count measurement. Let one mix unit contain $a_{A,16}$ native attention objects and $a_R$ recurrent checkpoints. When both counts are positive, the smaller of the attention and recurrent capacity ratios determines how many complete mix units can be stored. A zero-demand state type imposes no corresponding limit. The object counts refer to the stated storage scope and are not numbers of requests, layers, or seeds.

$$\widehat{C}_{D0}(M_A, M_R; a_{A,16}, a_R) = \min\left\{ \left\lfloor \frac{M_A}{16 a_{A,16}\,\text{KiB}} \right\rfloor, \left\lfloor \frac{M_R}{1557 a_R\,\text{KiB}} \right\rfloor \right\}. \tag{D2}$$

## A.2 Conditional Capacity of the Shared-Pool Designs

The same calculation can be stated for a shared-pool organization by testing integer repetitions of a chosen mix. For $a = (a_A, a_R)$, the variable $z$ counts repeated storage-mix units. The ideal S2 calculation uses $m = \lceil z a_R / 16 \rceil$ slabs. This gives a conditional upper model with ideal packing, not a guarantee that the serving runtime reaches that state count under its actual lifetime and allocation history.

$$\widehat{C}_c^{\text{ideal}}(M; a) = \max\left\{ z \in \mathbb{Z}_{\geq 0} : \widehat{B}_c^{\text{ideal}}(z a_A, z a_R) \leq M \right\}. \tag{A8}$$

**Table A-1. Storage for the same example state sets**

| Attention objects | Recurrent objects | Ideal S2 slabs | S0 MiB | S1 MiB | S2 MiB |
|---|---|---|---|---|---|
| 1 | 1 | 1 | 3.041016 | 2.500000 | 2.500000 |
| 16 | 16 | 1 | 48.656250 | 40.000000 | 32.500000 |
| 4 | 16 | 1 | 30.410156 | 34.000000 | 26.500000 |
| 0 | 16 | 1 | 24.328125 | 32.000000 | 24.500000 |

Arithmetic examples in one normalized storage scope. The recurrent-only row is not an attention-free serving configuration. With a 100-MiB budget and a=(1,1), the ideal capacities are 32, 40, and 49 mix units for S0, S1, and S2. These are not concurrent-request limits.

The examples show why a finer page is not automatically cheaper for every state mix. With only one convolution payload, a whole slab is still allocated, so S1 and S2 have the same cost in the first row. When sixteen convolution slots share a slab, S2 uses less space. By contrast, a recurrent-heavy mix can make S1 larger than S0 because its checkpoint allocation rises from 1,557 to 2,048 KiB. The capacity proxy therefore requires the mix and packing assumption to be stated.

## A.3 Encoding and Addressing a Convolution Slot

A packed convolution location identifies both a physical page and a slot within that page. For sixteen slots, the nonnegative location $\ell = 16p + s$ contains the actual page ID $p$ and slot offset $s$, with $0 \leq s < 16$. The address is computed from the convolution storage base using the 32-KiB slot stride. The page ID is not the position of a slab in a compact Python list, and a negative sentinel must not be decoded as a valid slot.

$$\begin{gathered} \ell = 16p + s, \quad 0 \leq s < 16, \\ \text{addr}_{\text{conv}}(\ell, o) = \text{base}_{\text{conv}} + \ell\, s_{\text{conv}} + o, \quad s_{\text{conv}} = 32\,\text{KiB}. \end{gathered} \tag{X2}$$

The stride and base must use the units of the particular access path. Byte offsets require byte strides; an element-indexed tensor view requires the corresponding element stride. The convolution base is local

to that storage and must not be substituted by the recurrent-state base. These address rules explain the representation and its guard conditions without claiming that a host mapping update proves completion of a device write.

### A.4 Compact Storage and Pool Flexibility Are Different Benefits

D0 can use fewer total bytes for a state set and still fail its per-pool limits. For a single recurrent object, D0 uses 1,557 KiB while S2 uses 2,048 KiB including one slab, leaving 491 KiB more allocation in S2. With sixteen objects in one slab, the extra allocation is 176 KiB, or 11 KiB per occupied slot. These differences describe packing cost, not the flexibility of sharing a total budget.

**Table A-2. Illustrative state set under a 100-MiB total budget**

| Quantity | D0 | S2 |
|---|---|---|
| Budget | 50 MiB attention + 50 MiB recurrent | 100 MiB shared |
| State set | 3840 native attention objects; 16 recurrent checkpoints | 120 attention objects; 16 recurrent checkpoints; 1 slab |
| Attention allocation | 60 MiB | 60 MiB |
| Recurrent and convolution allocation | 24.328125 MiB | 24.5 MiB |
| Total allocation | 84.328125 MiB | 84.5 MiB |
| Model feasibility | Fails attention limit: 60 > 50 MiB | Meets total limit: $84.5 \leq 100$ MiB |

Same represented state in the normalized geometry. This is a model example, not a measured maximum state set or a serving-performance prediction.

Table A-2 separates total compactness from feasibility under a fixed split. D0's total is slightly smaller, but its attention demand exceeds the attention allocation. S2's slightly larger total still fits the shared budget. This example illustrates the static dual-pool feasibility constraints in Section 3.5 without predicting a cache-hit or throughput gain for a particular execution.

## Appendix B. Detailed Experimental Methods and Measurement Scope

### B.1 Environment and the Time of Verification

Environment evidence distinguishes settings recorded by the run from values collected later from the maintained installation. The original campaign records identify the A100, BF16, tensor parallelism 1, and engine revision. CPU, container userland, host kernel, and library values were also collected in a September 17, 2026 snapshot. The operator reports that the environment was intentionally maintained, but the snapshot does not independently recover an absent package, kernel, or driver change history.

**Table B-1. Original campaign environment and later installation snapshot**

| Item | Reported value | Evidence scope |
|---|---|---|
| GPU; model arithmetic | A100-PCIE-40GB; BF16; TP=1 | Original run settings |
| Engine revision | ac4a65cc0797633b338135624a9e504c819762e5 | Original record; later same-HEAD clean-checkout report |
| vLLM version string | Run: 0.25.0+precompiled; later import: 0.25.0 | Distinct observations retained |
| CPU; RAM | 2 × AMD EPYC 9124; 32 physical cores / 64 threads; 2 NUMA nodes; 503 GiB RAM | Later snapshot; RAM is not the state-cache budget |
| Operating system | Debian 12 container userland; Linux 7.0.0-28-generic host kernel | Later snapshot; host and container are separate |

*Table B-1 (continued)*

| Item | Reported value | Evidence scope |
|---|---|---|
| Python; PyTorch; Triton | 3.11.2; import 2.11.0+cu130 / package 2.11.0; Triton 3.6.0 | Later import and package metadata |
| CUDA; driver | PyTorch build CUDA 13.0; driver 580.173.02 | Later snapshot; distinct version meanings |
| CUDA packages | Runtime 13.0.96; nvcc package 13.2.78 | Package values; system nvcc on PATH not established |
| cuDNN; NCCL | cuDNN API 9.19.0 (91900), package 9.19.0.56; NCCL 2.28.9 | API and package reports retained |
| Other serving packages | NumPy 2.3.5; Transformers 5.12.1; Tokenizers 0.22.2; Safetensors 0.8.0; FlashInfer 0.6.13 | Later snapshot |

This table reproduces the frozen evidence record, not a new check of current software. The later extension and policy campaigns have separate identity records.

Different version strings should not be collapsed into one apparently exact execution identity. The source-revision report explains the run/import vLLM distinction, but it does not make import metadata and package metadata identical. CUDA build support, installed runtime and compiler packages, and driver support also describe different things. The extension campaign retains its own model, engine, package, and actual-run manifest rather than inheriting every value from this earlier snapshot.

## B.2 Load Selection, Concurrency Limits, and Pinning Sensitivity

The load grids have documented selection and extension histories rather than one uniform preregistration history. Falcon's initial ladder follows an S0 pilot. The 4.5-request/s b108 point was added after the first three-seed results using a rule that multiplies the prior maximum by 1.5. The b63 cap ladder is 3, 5, 6, and 8; its largest value is the loosest tested cap, not necessarily the greatest pressure or effect. Reference runs use actual caps 63 and 109 from the existing 1.7-request/s cells, not unlimited concurrency or additional cap-sweep runs. Granite's 2/4/8-request/s ladder follows an S0-only pilot, but a pre-grid commit of the numeric ladder is not retained.

Sampled concurrency does not reveal every instantaneous scheduler decision. No extension poller sample has running=63, but one-second samples cannot exclude an intervening cap hit or directly count cap rejections. Separately, the low-load audit reports running maxima of 9–13 against configured caps 63/109. These observations should be stated at their sampling scope rather than generalized into a proof that concurrency limits never matter.

The pinning experiment changes a preemption-time control without disabling ordinary recurrent prefix caching. Roff disables extra checkpoint pinning at preemption. R0, R1, and R2 use distinct-requester thresholds K=3, 2, and 1. The rule is not a generic miss-demand counter: resident prefix-hit evidence and owner registration at preemption can contribute, while repeated references by the same requester are not counted again. Owner registration can allow an eligible K=1 candidate to pass immediately. The chosen rightmost boundary is tested against the threshold and protection budget; on resume, execution references are secured before extra pin references are released. Releasing a reference is not the same event as physically reassigning storage.

The pinning evidence uses a separate 96-cell robustness design. Two operating points, two organizations, four settings, and six paired seeds include 72 fresh cells and 24 compatible existing R1 cells. Cumulative pin events check whether the control is exercised; they do not measure simultaneous pinned occupancy. The two primary R0→R2 gap changes use a separate two-test Holm family. The eight small mean changes and the larger individual-seed maximum are both preserved in Appendix E.13, without an equivalence claim.

## B.3 Nominal Budgets and Integer Allocation

Planner matching preserves a nominal byte target while allowing integer allocation differences. Table B-2 gives the extension mapping. S0 and the smaller-page organizations require different overrides, while D0

uses a fixed split and integer native pages in each pool. Thus, varying D0's split can also change its rounded total. Neither the split sweep nor the complete-organization comparison is a perfectly equal-usable-byte experiment.

**Table B-2. Extension nominal budgets and planned allocation bytes**

| Budget | Nominal bytes | Overrides D0/S0/S1/S2 | D0 attention bytes | D0 recurrent bytes | D0 total bytes | S0 bytes | S1/S2 bytes |
|---|---|---|---|---|---|---|---|
| b63 | 101,154,816 | 63 / 63 / 193 / 193 | 53,346,304 | 46,236,672 | 99,582,976 | 100,445,184 | 101,187,584 |
| b83 | 133,267,456 | 83 / 84 / 254 / 254 | 70,270,976 | 62,180,352 | 132,451,328 | 133,926,912 | 133,169,152 |
| b108 | 173,408,256 | 108 / 109 / 331 / 331 | 91,439,104 | 81,312,768 | 172,751,872 | 173,786,112 | 173,539,328 |

Values are copied from the frozen budget table. These are planner/allocation-scope values, not a fresh dump of all runtime usable tensor storage.

**Table B-3. Model-specific native D0 split parameters**

| Model | Attention fraction | Recurrent fraction | Attention page bytes | Recurrent page bytes |
|---|---|---|---|---|
| Falcon | 0.527383 | 0.472617 | 16,384 | 1,594,368 |
| Granite | 0.199699 | 0.800301 | 32,768 | 1,074,688 |

The source selects the fixed fractions by prior model-specific worst stranded-byte screening, not by choosing the best subsequent serving result. Granite b160 has its own nominal mapping of 173,015,040 bytes and is not a Falcon b160 extrapolation.

## B.4 Counters, Time Windows, and Derived Service Metrics

Every work counter must be paired with the window and update rule that produced it. The earlier input accounting uses total recorded tokens H, record count K, and output-counter increment O_ctr over a common valid poller window. The difference H−O_ctr is a once-recorded input component, not all submitted or executed prefill. Throughput and latency instead refer to completed request episodes, while slab statistics are untimestamped release-event records. A same-run association does not make these observations the same time average.

The full-run, central-arrival, and full-poller views answer different questions. The serving span includes the initial growth and final drain of the finite workload. The earlier central view uses the 20th–80th percentiles of the common request-arrival sequence rather than taking the middle of each organization's completion duration. Comparing early and late running/waiting means in that window is a drift screen, not proof of steady state. The full one-second poller can start and stop outside either performance window. The confirmation's ledger episode is separately defined in Section 4.5 and must not be substituted for a request span.

Stranded-byte fraction measures a specific form of unused capacity in D0's split pools. It sums free bytes in one pool when the other pool is exhausted over scheduler steps and normalizes by configured pool bytes over those steps. This is step-weighted, not wall-time-weighted utilization or throughput loss. An absent corresponding field in a shared organization is not a measured zero.

Latency-qualified goodput in the auxiliary analysis uses thresholds derived from the same observed requests. Comparator-based and union-based empirical thresholds are applied back to those requests. This tests sensitivity to a latency definition; it does not validate compliance with an independently specified application SLO. Appendix E.8 therefore reports those quantities separately from plain completed-output throughput.

## B.5 Repetition Units and Original Statistical Families

The original five comparisons have a six- to nine-seed history that remains part of their interpretation. The b63 high-load comparison stays at six seeds, while the others reach nine through different extensions. Some additions follow a committed analysis revision; the burst and Granite triggering records were written after the extra runs. Therefore, a two-look adjustment does not make every extension prospectively identical. The five-test family also retains Granite after its results move to the appendix.

The original sign-flip tests require independent seed effects and sign symmetry under the null. The test enumerates sign combinations and compares the absolute mean with the observed absolute mean. Flipping a relative percentage is not equivalent to swapping the two throughputs, so a log-ratio sensitivity is reported separately. The two-look rule below adjusts the two opportunities to assess the same contrast; a second look that was not performed is assigned p9=1 under the recorded rule. Holm then adjusts decisions over the original five contrasts, not the measured effect sizes.

$$p_{\text{seq}} = \min\{1,\ 2\min(p_6, p_9)\}. \tag{B1}$$

The extension uses a fixed twelve-seed analysis of a partial factorial design. It spans three budgets, five D0 splits, and three sharing levels, but does not cross every budget, split, and sharing combination. Its main cells use 300 requests, 3 requests/s, cap 63, and seeds 42000–42011. The low-load extension uses 0.8 requests/s and only the first three of those seeds, so analysis keys retain load and study block to avoid replacing a main cell with a low-load cell.

The extension's 22-test family contains related contrasts rather than 22 independent discoveries. It includes six stepwise contrasts, three D0–S2 budget contrasts, two budget interactions, eight split contrasts, and three sharing-endpoint interactions. Displayed effects average paired relative changes or their percentage-point differences, but the tests use log ratios or log-ratio differences with all 4,096 sign assignments. Intermediate sharing and secondary counters are not promoted to primary tests after inspection. The fixed twelve-seed analysis has no six-to-nine continuation, but the pre-freeze smoke and order139 deviation remain disclosed. Rejections in this family, the original five, and the pinning family do not establish a single paper-wide 5% error rate.

## B.6 What the Artifact Checks Reproduce

Historical evidence checks establish different parts of the path from execution to result. The cap-sweep audit finds zero D0 pool rejections inside its swept range, not a disabled or free allocation check. Granite's earlier four-arm comparison explains model dependence but is not merged with later six- or nine-seed runs. The extension closeout connects released tables, accounting, and retained records while keeping raw loss and rerun history visible. Later handoff status is described separately in B.12 and F.5.

The input-accounting and work-timing reanalyses provide stronger raw-to-result checks within their supplied scopes. The source reports reproducing the earlier accounting from 36 runs' request, poller, and server logs and checking four counter-path source files. The separate work-timing analysis regenerates thirteen final CSVs byte-for-byte from 131 inputs, with an independent parser checking 939,617 trace records. Those checks support the correspondence between recorded data and results. They do not independently certify every instrumentation emitter, wrapper, collector, or the complete historical execution binary.

## B.7 Applying the Storage Model to Observations

Applying a storage equation first requires fixing model geometry, object identity, and accounting scope. A comparison of the same state set under two representations isolates a calculated representation cost. A comparison of the different sets retained in two executions instead describes system behavior. Even when allocator observations can be inserted into an equation, the result is an accounting reconstruction, not a measured maximum capacity unless saturation and feasible state mix have also been established.

Work and time should be connected only over compatible runs, seeds, cohorts, and windows. The once-recorded input counter is not submitted or completed replay. A convolution live/page dump can compare representations of that particular set, but does not measure a different S1 execution or all hybrid-state capacity. Recurrent free-byte observations also require layer/group normalization before comparison with a planner budget. Sections 3.3--3.5 distinguish representation cost from static-pool feasibility and do not directly predict cache hits or throughput.

## B.8 Submitted-Work and Event-Time Diagnostic

The submitted-work diagnostic is a separate three-seed study of the same-run relation between work, event time, and performance. It reports Falcon, A100, BF16, and TP 1 but distinguishes base engine ac4a65cc07 from instrumented engine 8adbe9953e. Harness revisions and missing per-run loaded-source hashes prevent a claim of identical historical binaries. Its three operating points each cross D0/S2 and reference/traced modes, giving 36 workload-complete runs. Neither earlier six/nine-seed runs nor the extension's cap 63 twelve-seed results are additional repetitions of this diagnostic.

**Table B-4. Operating points of the submitted-work diagnostic**

| Point | Budget | Requests/s | Cap | Conditions | Seeds | Runs |
|---|---|---|---|---|---|---|
| L | b63 | 3.0 | 63 | D0/S2 × REF/TRACE | 0–2 | 12 |
| M | b108 | 3.0 | 109 | D0/S2 × REF/TRACE | 0–2 | 12 |
| H | b108 | 4.5 | 109 | D0/S2 × REF/TRACE | 0–2 | 12 |

Eighteen runs are traced. REF/TRACE here concerns work and timing instrumentation, not the state-history OFF/ON pair or a correctness recomputation reference.

The diagnostic counts valid submitted tokens and classifies work from the pre-update schedule. An original submit-side subclassification was emitted after token positions changed; linking to the pre-update schedule corrects that classification. The valid-token total was separately retained and is not a complete position-level division into new work and replay. CUDA-event elapsed times are summed over the same token-bearing forwards. Zero-token teardown records are distinguished from missing timing for actual work, and the sum is not GPU kernel-active union or CPU compute time.

Pool shortage is assessed by need exceeding free capacity, not by a zero free count alone. A successful check with need=free=0 does not become a failure. Recurrent byte values require the reported layer weighting to be reconciled with the planner scope. Performance uses the mean seedwise relative effect, whereas submitted-work and event-time ratios use pooled TRACE totals. Reference-to-trace changes use only their common valid seeds, not a mix of incompatible observations.

The full traced record is the primary diagnostic view, with realized-arrival central windows and probe effects reported separately. Those central windows are reconstructed for each run and are not the original shared frozen-arrival window. The source-input reproduction closes the checked transformation into result tables, while unresolved replay and missing emitter/collector source remain different limitations. Gate-off smoke checks establish recorded counts and configuration, not exact output-token or state-byte equality.

## B.9 Executed Kernel Paths

Executed flags, rather than later defaults, identify the S1 and S2 kernel path in the performance comparisons. The path audit reports 183 boot logs: 120 extension and 63 earlier runs. A later index ties six representative b83 sharing-level runs to retained startup records and notes that the zero-sharing rows are reused by the counter and sharing analyses. This supplementary index supports those representative links; its delivery is not a new independent inspection of all 183 logs.

The measured S1 and S2 runs explicitly enable paged recurrent execution. An omitted flag in a later default configuration is a different condition. Direct per-token state access also differs from gathering a

starting state during input preparation or recovery. The source audit attributes the additional convolution-slot flag to S2, but a one-flag configuration difference still includes the implementation's addressing, protection, and reclaim effects. It is not a cost-free capacity-only intervention.

## B.10 Byte Scope and Unrecorded Usable Memory

The nominal-budget constant does not change the logical checkpoint interval. The handoff connects 1,605,632 bytes to a normalization calculation of 1,568 tokens times 1,024 bytes/token. That explains the budget-label mapping; it does not replace D0's 1,792-token interval or the shared designs' 512-token boundary. Representative recurrent and convolution payloads, planned allocation, actual tensor storage, and runtime usable memory remain distinct quantities.

The reported b63 D0/S2 byte difference is a planner-scope comparison rather than full runtime equality. D0's logged pool plans and S2's page product give about $-1.5858\%$. The handoff's actual-tensor column connects an existing allocation audit and source interpretation, not a newly supplied per-layer tensor-storage dump. Usable runtime fields remain unrecorded at that scope. Thus, neither planner matching nor those columns prove equality of every physical or usable allocation throughout a run.

## B.11 Workload, Protection Settings, and Model Identity

Model and workload identity is tied to the campaign that recorded it. The handoff separates earlier performance, extension, independent confirmation, short probes, and state-history runs. The reported model name is Falcon-H1-7B-Instruct, but the extension is the group explicitly linked to a frozen model snapshot and tokenizer hash. Reusing a local path in another campaign does not retrospectively identify its snapshot. Request counts, input rules, and protection thresholds likewise remain campaign-specific.

The historical protection-contract typo is corrected in a derivative record without overwriting the frozen original. Two historical JSON strings say LF_K=1 even though the original and extension runs use K=2. The supplied correction changes only those strings, retaining the source SHA and a distinct derivative SHA. The extension manifest's 348 LF-K values and the runner injection support K=2. The later admission confirmation, short probes, and state trace instead use K=1. This documentation correction changes neither measurements nor the claim that different K values can have different effects.

## B.12 Statistical Artifacts, Display Intervals, and Adoption Sensitivity

The statistical handoff connects the reported effects to their original families instead of introducing new tests. Its 81-row map and nine-row sensitivity record cover extension tests, admission confirmation, simultaneous bounds, and earlier contrasts. The extension's 22-test analysis has no confidence intervals; the observed range and sign counts in Table B-5 are not confidence intervals. Confirmation display intervals use the original 10,000 seed-level resamples and are not interchangeable with either Holm decisions or the 36 simultaneous upper bounds.

**Table B-5. Stepwise extension effects and observed seed variation**

| Budget | Contrast | n | Mean change | Observed min–max | Positive / negative | Holm p | CI |
|---|---|---|---|---|---|---|---|
| b63 | S1/S0 | 12 | +51.784% | +45.840% – +56.508% | 12 / 0 | 0.01074219 | Not supplied by original analysis |
| b83 | S1/S0 | 12 | +47.568% | +41.152% – +56.104% | 12 / 0 | 0.01074219 | Not supplied by original analysis |
| b108 | S1/S0 | 12 | +56.235% | +37.395% – +73.191% | 12 / 0 | 0.01074219 | Not supplied by original analysis |
| b63 | S2/S1 | 12 | +8.929% | +5.939% – +11.589% | 12 / 0 | 0.01074219 | Not supplied by original analysis |

*Table B-5 (continued)*

| Budget | Contrast | n | Mean change | Observed min–max | Positive / negative | Holm p | CI |
|---|---|---|---|---|---|---|---|
| b83 | S2/S1 | 12 | +12.628% | +6.504% – +28.120% | 12 / 0 | 0.01074219 | Not supplied by original analysis |
| b108 | S2/S1 | 12 | +9.972% | -0.058% – +29.926% | 10 / 2 | 0.03662109 | Not supplied by original analysis |

Original 22-test family; ranges are across twelve seed effects. The current adoption history is retained.

The seed-exclusion sensitivity retains both the original result and the affected contrast's changed sample size. The two b108 stepwise contrasts include the adopted order139 rerun in their n=12 main analysis. Excluding seed 42004 gives n=11 only for the affected contrasts while retaining the complete 22-test correction family. The resulting averages and adjusted values match the recorded sensitivity, but do not recover the lost first-attempt raw or remove protocol deviation.

**Table B-6. b108 main and seed-exclusion analyses**

| Contrast | Scenario | n | Mean change | Holm p |
|---|---|---|---|---|
| S1/S0 | Current adoption | 12 | +56.235% | 0.01074219 |
| S1/S0 | Exclude 42004 | 11 | +55.774% | 0.01074219 |
| S2/S1 | Current adoption | 12 | +9.972% | 0.03662109 |
| S2/S1 | Exclude 42004 | 11 | +10.879% | 0.03417969 |

Post hoc sensitivity, not a replacement of the main estimand or an equivalence analysis. The invalid-adoption scenario is retained in Appendix E.14.

Individual display intervals can disagree with a family-adjusted rejection decision without a calculation error. Table B-7 includes thirteen supplied display intervals, not invented intervals for the two self-comparisons or two cross-budget contrasts. In particular, b83 at zero headroom has a display interval just below zero, but adjusted p=0.30761719 and no rejection. The budget-endpoint interaction also retains its original adjusted p=0.13671875. Neither a different effect display nor a post hoc interval replaces those decisions.

**Table B-7. Admission-confirmation display intervals and adjusted decisions**

| Legacy contrast ID | Contrast | n | Mean change | Individual 95% interval | Holm p | Reject |
|---|---|---|---|---|---|---|
| P0-b63 | D0(0.00)/S2(0.00) @ b63 | 12 | +4.431% | [+2.911%, +5.944%] | 0.00830078 | True |
| P19-b63 | D0(0.19)/S2(0.19) @ b63 | 12 | -8.420% | [-9.836%, -6.756%] | 0.00830078 | True |
| PT-b63 | D0(0.00)/S2(0.27) @ b63 | 12 | -8.013% | [-9.863%, -6.147%] | 0.00830078 | True |
| IS-b63 | S2(0.27)/S2(0) @ b63 | 12 | +13.613% | [+11.658%, +15.393%] | 0.00830078 | True |
| P0-b83 | D0(0.00)/S2(0.00) @ b83 | 12 | -3.269% | [-7.084%, -0.023%] | 0.30761719 | False |
| P19-b83 | D0(0.19)/S2(0.19) @ b83 | 12 | -16.348% | [-17.783%, -14.731%] | 0.00830078 | True |
| PT-b83 | D0(0.10)/S2(0.35) @ b83 | 12 | -16.218% | [-17.395%, -15.095%] | 0.00830078 | True |
| ID-b83 | D0(0.10)/D0(0) @ b83 | 12 | +2.857% | [+2.559%, +3.139%] | 0.00830078 | True |
| IS-b83 | S2(0.35)/S2(0) @ b83 | 12 | +18.808% | [+14.047%, +23.214%] | 0.00830078 | True |
| P0-b108 | D0(0.00)/S2(0.00) @ b108 | 12 | -14.070% | [-16.428%, -11.842%] | 0.00830078 | True |

*Table B-7 (continued)*

| Legacy contrast ID | Contrast | n | Mean change | Individual 95% interval | Holm p | Reject |
|---|---|---|---|---|---|---|
| P19-b108 | D0(0.19)/S2(0.19) @ b108 | 12 | -11.993% | [-14.429%, -9.644%] | 0.00830078 | True |
| PT-b108 | D0(0.10)/S2(0.00) @ b108 | 12 | -11.936% | [-14.211%, -9.697%] | 0.00830078 | True |
| ID-b108 | D0(0.10)/D0(0) @ b108 | 12 | +2.501% | [+1.789%, +3.231%] | 0.00830078 | True |

Copied from the original 17-comparison family. These are display intervals, not the 36-family simultaneous bounds. True/False retain the recorded rejection decision.

Absolute confirmation latency is summarized at the run level before averaging seeds. Table B-8 groups the 360 adopted runs into thirty organization–budget–headroom cells. Each latency entry is the average of twelve run-specific p99 values, not a pooled percentile. The final column concerns request-level average output-token interval, not the distribution of every token gap. This is a source-based regrouping of existing measurements, with no new samples, intervals, or statistical decisions.

**Table B-8. Absolute request-rate and latency summaries in admission confirmation**

| Budget | Org. | f | Runs | Requests/s | TTFT p99 (s) | E2E p99 (s) | Request TPOT p99 (s/token) |
|---|---|---|---|---|---|---|---|
| b63 | D0 | 0.00 | 12 | 1.9182 | 49.7276 | 62.6483 | 0.057875 |
| b63 | D0 | 0.10 | 12 | 1.9249 | 49.6962 | 61.3189 | 0.049890 |
| b63 | D0 | 0.19 | 12 | 1.8764 | 53.5786 | 64.4228 | 0.050679 |
| b63 | D0 | 0.27 | 12 | 1.8149 | 59.0705 | 69.0041 | 0.044725 |
| b63 | D0 | 0.35 | 12 | 1.7103 | 69.4591 | 78.1230 | 0.043654 |
| b63 | S2 | 0.00 | 12 | 1.8381 | 56.7566 | 68.3108 | 0.053367 |
| b63 | S2 | 0.10 | 12 | 1.9704 | 46.0887 | 57.5037 | 0.046080 |
| b63 | S2 | 0.19 | 12 | 2.0506 | 40.6063 | 51.2309 | 0.040384 |
| b63 | S2 | 0.27 | 12 | 2.0877 | 38.5759 | 48.7259 | 0.036245 |
| b63 | S2 | 0.35 | 12 | 2.0430 | 41.4172 | 50.6719 | 0.033737 |
| b83 | D0 | 0.00 | 12 | 2.1671 | 31.1423 | 50.3830 | 0.071288 |
| b83 | D0 | 0.10 | 12 | 2.2290 | 27.4726 | 45.2691 | 0.059008 |
| b83 | D0 | 0.19 | 12 | 2.2074 | 29.1234 | 45.0813 | 0.056697 |
| b83 | D0 | 0.27 | 12 | 2.1352 | 34.0909 | 48.1545 | 0.052012 |
| b83 | D0 | 0.35 | 12 | 2.0261 | 41.9435 | 53.9381 | 0.049659 |
| b83 | S2 | 0.00 | 12 | 2.2514 | 26.6408 | 43.3550 | 0.059271 |
| b83 | S2 | 0.10 | 12 | 2.4400 | 16.7495 | 34.1919 | 0.051186 |
| b83 | S2 | 0.19 | 12 | 2.6413 | 8.0169 | 26.8011 | 0.043908 |
| b83 | S2 | 0.27 | 12 | 2.6961 | 6.4804 | 24.1752 | 0.038687 |
| b83 | S2 | 0.35 | 12 | 2.6620 | 8.1239 | 24.0176 | 0.035094 |
| b108 | D0 | 0.00 | 12 | 2.4063 | 16.7541 | 46.1061 | 0.085953 |
| b108 | D0 | 0.10 | 12 | 2.4665 | 14.5379 | 41.1534 | 0.073607 |
| b108 | D0 | 0.19 | 12 | 2.4636 | 14.9338 | 39.1464 | 0.064354 |
| b108 | D0 | 0.27 | 12 | 2.4126 | 17.3618 | 39.0158 | 0.061024 |
| b108 | D0 | 0.35 | 12 | 2.3280 | 21.9305 | 40.0692 | 0.055488 |
| b108 | S2 | 0.00 | 12 | 2.8055 | 0.7565 | 19.6847 | 0.040454 |
| b108 | S2 | 0.10 | 12 | 2.8056 | 0.7485 | 19.3214 | 0.036693 |
| b108 | S2 | 0.19 | 12 | 2.8043 | 0.7929 | 19.6345 | 0.037564 |
| b108 | S2 | 0.27 | 12 | 2.8049 | 0.8604 | 19.6519 | 0.037460 |

*Table B-8 (continued)*

| Budget | Org. | f | Runs | Requests/s | TTFT p99 (s) | E2E p99 (s) | Request TPOT p99 (s/token) |
|---|---|---|---|---|---|---|---|
| b108 | S2 | 0.35 | 12 | 2.8047 | 1.1658 | 19.4673 | 0.036378 |

Arithmetic means over twelve runs per cell. All source precision is retained; displayed p99 values and throughput timebases are not redefined.

# Appendix C. Experiment Catalogue, Supporting Comparisons, and Terminology

## C.1 Experiment Roles and Supporting Results

The experiment catalogue distinguishes paper questions from historical experiment identifiers. Current Q1–Q9 are the organizing questions in Table 2. Some source rows have older numbers with different meanings; the catalogue below names their role and retains the original identifier only for traceability. An analysis of an existing run is not an additional GPU repetition. In particular, the state trace and joint-depth reanalysis share data, and the trace-fidelity check uses the same paired modes rather than another 24 runs.

**Table C-1. Catalogue of main and supporting experiments**

| Paper role | Source experiment IDs | Design or observation | Where the evidence is used |
|---|---|---|---|
| Stepwise performance, Q1 | 1a; 1b | Twelve-seed main comparison; separate three-seed low load | Section 5.1; Appendix E.1 |
| Work and state, Q2 | 2a; 2b; 2c | Two earlier anchors; zero-sharing extension condition | Section 5.2; Appendix E.1 |
| Sharing, Q3 | 3a | b83 sharing fractions 0/0.5/1 | Section 5.2; Appendix E.6 |
| Budget, Q4 | 4a plus reused 1a | D0/S2; existing S0/S1/S2 results on the common budget axis | Section 5.3; no new four-arm run |
| Arrival rate, Q5 | 4b; 4c | b63/b108 curves and separate enlarged seed sets | Section 5.3; Appendix E.2 |
| Concurrency limits | 4d; 4e | b63 caps 3/5/6/8 and b108 caps 5/8/10/13; existing references | Appendix E.3 |
| Arrival bursts | 4f | Poisson versus five seconds on, fifteen off at mean 1.7 requests/s | Appendix E.4 |
| D0 pool split | 4g | Five attention fractions at b63/b108 | Appendix E.5 |
| Preemption-time pinning | 4h | Roff; K3/2/1; two points; six seeds; 96 cells | Appendix E.13; not current Q9 |
| Admission development | 4i; 4j; 4k | Initial b63/b83/b108 five-value search, seeds 76200–76202 | Selection history, not confirmation data |
| Admission extended development | 4l | b83 f=0/0.35/0.43/0.50/0.60, seeds 76300–76302 | Separate search because the baseline changed |
| Admission confirmation, Q6 | 4m; 4n; 4o | Three budgets × two organizations × five f values × twelve seeds | Section 5.4; Appendix B.12 |
| Admission semantics | 5a; 5b | Code-path and allocation checks | Methods; not separate performance effects |
| Short-load probes | 5c; 5d; 5e | Forty-request high-rate diagnostics | Auxiliary only; distinct seeds across f |
| State availability, Q7 | A3a; JR-1 | Native D0/S2 trace and offline joint-depth reanalysis | Section 5.5; Appendices C and F |
| Trace functionality and fidelity | A2a; A3b | One-seed precursor; three-seed paired final check | Appendix D; final check shares the trace campaign |

*Table C-1 (continued)*

| Paper role | Source experiment IDs | Design or observation | Where the evidence is used |
|---|---|---|---|
| Boundary-matched reuse/work | 6a; 6b | D0-M512/S2, low/high usable-budget conditions, six seeds | Auxiliary input-work accounting |
| Boundary-matched reuse isolation | 7a | D0-M512/S2, f0/0.19 × allowed/ isolated, three seeds | Fixed-order pilot; not current paper 7a |
| Submitted work and scheduling | 7b; 8a | Eight conditions, three seeds; final scheduler work and step intervals | No unrecorded CPU/GPU durations inferred |
| Timing instrumentation | 8b | D0-M512/S2 × f0/0.19 × M0/MH/MT × nine seeds | Separate 108-run time study |
| Fixed-work width | 9a | Sixteen decode requests once versus eight twice | Local D0-M512 cost, not current paper 9a |
| Fixed-work mixing | 9b | Fifteen decode plus one input chunk; mixed versus separated | CUDA-event cost includes execution-mode differences |
| Live separation policy | 10a | Baseline versus input/decode separation at f0/0.19 | Intervention fails empty-check-step tolerance |
| Checkpoint lifetime, Q8 | RR-1 Phase B | S2 versus S2+Retention at 254 pages | Section 5.5; exploratory three-seed intervention |
| Retention confirmation, Q9 | P1 | S2/S2+Retention × shared/isolated; fresh six seeds | Section 5.5; primary ledger K_ret |
| Continuation checks | C1R-CONT chain | Clean references, repeatability, isolation, fidelity | Appendix D; not a performance-pass inference |

Source IDs are not new paper question numbers. All original source tables remain in the accompanying frozen-source evidence catalogue, including values not selected for repeated typesetting here.

Admission development chooses settings before independent confirmation rather than reselecting the highest later mean. The initial three-budget search uses seeds 76200–76202 and f=0/0.10/0.19/0.27/0.35. It selects D0/S2 values 0/0.27 at b63 and 0.10/0 at b108. The separate b83 extension uses seeds 76300–76302 and f=0/0.35/0.43/0.50/0.60, selecting S2=0.35 from candidates within 1% of the highest mean. Its zero-headroom S2 baseline differs from the first search by 11.01%, above the recorded 5% merging tolerance, so those curves are not pooled. Missing intermediate absolute S2 values in the first b83 search are not reconstructed from rounded percentages.

The final state-history analysis reports modelled entry depth and measured adopted depth separately. All 3,714 admission episodes bind adopted depth to request and preemption count. The block-ID entry reconstruction changes 123 episode values and yields 669 zero-to-zero, 2,188 positive-equal, 432 newly reusable, 40 deeper, and 385 reduced-depth episodes. Entry depth remains an offline model, not engine ground truth at queue entry. The corrected scalar replay and final-depth lineage each match 3,714/3,714 endpoints, but do not reconstruct every producer or the full cache map.

**Table C-2. Initial-admission state-depth classes in the final trace analysis**

| Organization | f | Zero→zero | Positive equal | Zero→positive | Deeper | Shallower |
|---|---|---|---|---|---|---|
| D0 | 0 | 0.137 | 0.644 | 0.120 | 0.000 | 0.099 |
| D0 | 0.19 | 0.080 | 0.766 | 0.080 | 0.000 | 0.074 |
| S2 | 0 | 0.362 | 0.332 | 0.153 | 0.019 | 0.133 |
| S2 | 0.19 | 0.162 | 0.564 | 0.127 | 0.026 | 0.121 |

Three-seed means for initial episodes, at the source display precision. These fractions are not pooled over both headroom settings. Positive-equal depth does not prove uninterrupted survival of the same physical generation.

The corrected lineage closes a descriptive inclusion relation, not the complete headroom-to-throughput causal chain. It records 7,476 depth-decreasing loss events and 19,918 depth-increasing publication events. An earlier-sibling same-step relation reconciles 288 events in three episodes, while 40,787 same-step publications remain ambiguous as to producer identity. Every one of the 385 final reduced-depth episodes in-

cludes at least one other-request allocation-associated depth-decreasing loss. That inclusion does not make the event the unique cause or assign a mediation percentage. Request-ID-only overwrite results from the superseded analyzer are not retained as current evidence.

**Table C-3. Throughput probe-effect containment for the state trace**

| Organization | f | Mean 100 ln(ON/OFF) | 90% paired t interval | [−3, +3] containment |
|---|---|---|---|---|
| D0 | 0 | −0.361 | [−1.955, +1.233] | PASS |
| D0 | 0.19 | −0.280 | [−0.395, −0.164] | PASS |
| S2 | 0 | +0.431 | [−1.886, +2.749] | PASS |
| S2 | 0.19 | −0.779 | [−1.069, −0.489] | PASS |

Exactly three paired seeds, df=2, in the same 24-run collection. This is margin containment, not zero effect or equality of every internal state distribution.

The boundary-matched auxiliary studies examine a different D0 variant from the native organization in the main comparison. D0-M512 uses 512-token attention and recurrent boundaries. With six seeds at 3 requests/s, cap 63, and 300 requests, increasing f from 0 to.19 increases adopted-prefix share and reduces submitted input in the reported low and high usable-budget settings. At the low setting, S2's initial submitted input falls from 2,203.058 to 1,514.702 tokens/request; at the high setting it falls from 2,180.018 to 1,478.009. These are final submitted-work quantities in that study, not the once-recorded counter or a measurement on native D0.

**Table C-4. Boundary-matched auxiliary initial and resumed input work**

| Budget setting | Organization | f | Adopted prefix/ input (%) | Initial tok/req | Resume tok/req | Total tok/req |
|---|---|---|---|---|---|---|
| Low | D0-M512 | 0 | 39.802 | 2,020.162 | 42.156 | 2,062.318 |
| Low | D0-M512 | 0.19 | 46.850 | 1,788.338 | 0.000 | 1,788.338 |
| Low | S2 | 0 | 34.527 | 2,203.058 | 0.000 | 2,203.058 |
| Low | S2 | 0.19 | 54.982 | 1,514.702 | 0.000 | 1,514.702 |
| High | D0-M512 | 0 | 41.203 | 1,971.764 | 43.802 | 2,015.566 |
| High | D0-M512 | 0.19 | 48.600 | 1,729.458 | 0.000 | 1,729.458 |
| High | S2 | 0 | 35.212 | 2,180.018 | 0.158 | 2,180.176 |
| High | S2 | 0.19 | 56.073 | 1,478.009 | 0.000 | 1,478.009 |

Six-seed means, seeds 70000–70005, separate usable-budget study. Resume includes the defined catch-up category and is not automatically pure replay.

The earlier isolation pilot shows reuse-dependent admission effects but retains its fixed execution-order limitation. At 3 requests/s, cap 63, and seeds 71000–71002, S2's f=0.19 effect is +10.41% with reuse allowed and −11.50% with cross-request reuse blocked. D0-M512's corresponding effects are −1.25% and −6.39%. Twelve isolated performance cells have zero prefix-hit tokens; a separate two-run functionality check records 114 versus 0 actual adoptions and is not added to the performance sample. The source's log-scale interaction is not a percentage-point subtraction or the later balanced retention confirmation.

Submitted-work and scheduling records explain what changed without supplying unrecorded durations. In the eight-condition three-seed follow-up, isolated S2 has approximately 1.089 million submitted tokens at both headroom values, while shared S2 falls from.772 to.534 million. Average finally scheduled requests and host step spacing fall, but the number of work-bearing steps rises. That record does not contain request-level admission, CPU preparation, or GPU duration boundaries, so step spacing cannot fill those fields. A later 108-run M0/MH/MT study supplies separate timing evidence and reports a positive S2 headroom effect; its timing-policy fidelity remains limited, so the elapsed components are not converted into fractions of the total gain.

Fixed-work experiments show local batching costs without identifying a serving-wide mechanism. A valid D0-M512 fixture at context 3072 compares sixteen decode requests once with eight requests twice; the wider operation has lower per-token device cost. Fixtures exceeding the recorded 1% disagreement tolerance are excluded. A second fixture mixes fifteen decode requests with one input chunk. Separating the two operations has about 7 ms lower summed event time for 512 input tokens and about 3.6 ms lower time for 1536 tokens, with the recorded 0/16 logical-result differences. Mixed input uses ordinary execution, while separated decode uses full graph replay, so this is not a pure interference-cost measurement.

The live separation intervention does not establish that mixed execution causes the headroom benefit. It removes mixed work-bearing steps but increases empty admission-check steps beyond the preset 5% tolerance in all twelve separated runs. Throughput declines in all four organization–headroom conditions, and the reported TTFT/E2E changes are positive. The baseline lacks the same empty-check diagnostic, and device execution-mode or empty-step costs are not recorded. Thus, the failed intervention validity condition prevents a pure causal interpretation; it is neither proof that mixing is harmless nor a successful serving optimization.

The retained-state experiments use their own page budgets, seeds, and evidence hierarchy. Current Q8 refers to the three-seed same-budget checkpoint-lifetime intervention. Current Q9 refers to the fresh six-seed balanced performance confirmation with ledger K_ret primary, not the old fixed-work experiment called 9a. The continuation chain supplies a different kind of evidence and remains unresolved as described in Appendix D. The catalogue keeps these distinctions even where historical source labels overlap.

### C.2 Expanded Original Results and Data Lineage

Expanded tables use existing result inputs at their original aggregation level. The load and cap curves come from the earlier figure-mean CSVs and run ledgers. The stepwise secondary latency table groups the adopted 348-row extension input by its original conditions and twelve seeds. Expanding those rows does not add a run or a new significance test. Missing values remain distinguished from unrecorded and inapplicable quantities.

The manuscript catalogue and the full source catalogue serve different reading needs. Main results and selected detailed tables are typeset in English here, while the accompanying authoring evidence catalogue preserves every table element from the frozen Logic Flow, its original text, source anchor or ordinal, and a content hash. That archive includes historical and editorial records as well as current scientific tables; it is not presented as 209 independently collected experiments or 209 mandatory submission tables. Appendix F distinguishes manuscript content from the working evidence and public-release package.

### C.3 Definitions for the Supporting Measurements

Headroom, submitted work, and scheduler counts use different denominators and should be read separately. The headroom reserve is an admission requirement, not memory newly allocated to the admitted request; available blocks can include reclaimable cached blocks. Submitted work is the final scheduled token assignment after same-step rollback, not completed FLOPs. INITIAL and RESUME are catch-up categories separated by preemption history, while NORMAL_DECODE is scheduled decode work and need not equal client output tokens. Work-bearing step counts and average selected request counts are not kernel counts or time-weighted occupancy.

Instrumentation mode labels describe extra recording rather than a different compute device. M0 has no added timing, MH adds host intervals, and MT adds host timing plus deterministic 1/16 CUDA-event sampling. All still run the model on the GPU. Host and device spans can overlap and cannot be summed into E2E. A fixed-work separated total is the sum of its component medians under the recorded execution formats, not a sum of all serving delays. These modes also differ from the state-history performance/trace pair.

State-history terms identify a depth and an event meaning, not complete physical provenance. Entry depth is reconstructed at the waiting-entry cut; adopted depth is the actual admission coordinate. Positive-equal

depth does not prove continuous survival of the same generation. Publication means making a cache key available, not GPU-write completion; a loss is a recorded loss of supply, not every possible eviction. D0's 16/1792 and S2's 512/512 reconstruction rules must match their respective geometry. The final corrected analysis supersedes older progress reports without retroactively converting an unrecorded producer or full-map snapshot into known data.

## Appendix D. Implementation, Trace Integrity, and Correctness Evidence

### D.1 Source-Audit Scope

A source audit identifies the revision and execution configuration it actually inspected. The later September 23 audit concerns engine 51cc2d9e95 and its then-current D0-M512/S2 setup. Its S0/S1 observations are code-level scope, not new measured performance runs. The earlier S1/S2 performance path must instead be tied to its explicit paged-kernel flags and startup records in Appendix B.9. A later default or dispatch change is not evidence that the historical measured path used the same default.

### D.2 Safety Invariants and the Evidence Needed

The design requires a complete reusable boundary, protected accesses, ordered invalidation, and explicit absent-state handling. Table D-1 states these as conceptual invariants, not as invented names of four runtime assertions. Each needs its own implementation or execution evidence. Passing an allocator fixture cannot by itself certify every possible retained continuation.

**Table D-1. Design invariants and verification boundaries**

| Invariant | Required behavior | Evidence boundary |
|---|---|---|
| Complete checkpoint | All necessary recurrent members and convolution state are valid at the selected boundary | A partial member hit is not accepted as a complete logical hit |
| Protected access | Storage accessed by an active request has the required references | Representative protection alone is insufficient if other storage is used |
| Invalidate before reuse | Affected cached composites become invalid before storage is reassigned | Last-reference release and physical reuse are distinct events |
| Valid location or absence | Zero can be a valid location; a negative sentinel represents absence | Guarded missing state is not a valid zero-filled checkpoint by assumption |

Conceptual audit categories from the frozen design appendix, not an assertion of universal end-to-end correctness.

### D.3 Implemented Paths and Remaining Scope

The composite mapping expands a logical representative into the storage that must be accessed and protected. The audit describes member expansion and touching, explicit paged recurrent access, guarded missing indices, and convolution-slot dependencies. Persistent mapping buffers and their device updates belong to the particular runner path. These source observations do not by themselves establish a universally synchronized host/device table or an implicit full-checkpoint copy on adoption.

Slab reclaim couples the validity of checkpoints that occupy the same backing page. Active-slot references protect the slab, whereas inactive cached entries can remain valid until reassignment. Reclaim must invalidate every affected cached composite before reuse. This implementation evidence supports the lifetime rule, but a slab event count is not a duration or a complete instantaneous byte-accounting record. The 512-token boundary and page geometry are evaluated settings, not proven optima.

## D.4 Collection Completion and Shutdown Integrity

A workload can finish even when the trace tail is incomplete. The initial D0-b83-f.19 traced attempt and its retry both finish 300/300 requests but lack the required end record. A subsequent PID-only zero-timeout probe also lacks it. A single ten-second-grace probe produces an end record at sequence 62041 and is a diagnostic boot, not one of the final 24 scientific runs. These histories prevent request completion from being used as a substitute for writer-tail integrity.

The final collection uses the same shutdown rule across all 24 planned cells. The preparation checks report 69/69 CPU tests, and the final new-root collection completes 24/24 without a halt or retry. The twelve traced runs have the required end record, no sequence gap, no dropped or write-error count, and 300/300 requests. The twelve performance runs are not expected to emit that trace end marker. The 3,714 admission episodes are then analyzed at their recorded scope. A fallback before the required end is a failure; cleanup after the end does not prove every unrelated worker action is covered.

**Table D-2. Final trace collection and its separate precursor checks**

| Evidence | Outcome | Use |
|---|---|---|
| Initial completed workload without end marker | 300/300 requests; required tail absent | Excluded from formal trace collection |
| Ten-second shutdown probe | One boot; end sequence 62041 | Diagnostic only |
| Final 24-run collection | 24/24 accepted; halt 0; retry 0 | Performance 12 + trace 12 |
| Final trace records | End present; gap 0; dropped 0; write errors 0 | Recorded trace-integrity scope |
| Throughput probe effect | 4/4 interval-containment checks pass | Same 24 runs; not exact-zero overhead |
| Admission endpoint binding | 3714/3714 | Per-episode analysis, not 3714 independent workload repeats |

Precursors are not added to the final scientific sample. Trace-tail completion and output correctness remain separate checks.

## D.5 Numerical and Observable-Continuation Checks

The numerical checks cover different units and do not combine into one universal correctness result. S1 unit tests use batch sizes 8,32,53 with BF16 outputs and FP32 recurrent state, and report exact tensor comparisons for the checked output and destination. S2 has corresponding slot tests. Forced-reclaim and deeper-prefix tests use other workloads and references. Table D-3 retains those scopes instead of replacing exact checks with a successful process exit.

**Table D-3. Selected implementation and numerical validation scopes**

| Check | Recorded result or scope | Limit |
|---|---|---|
| S1 paged-kernel unit test | Batches 8/32/53; exact output/destination tensor comparisons | BF16 output with FP32 state; not every serving dtype/path |
| S2 convolution-slot unit test | Checked output and state comparison reported | Return code alone is not the numerical oracle |
| Granite forced reclaim | 28 planned conditions including 4 references;24 comparison tests | Separate model and greedy shared-prefix workload |
| Falcon pre-logit checks | Depth 600–1800:18/18 hashes match; depth 2400:6/6 mismatch and 4/6 token mismatch | Does not support uniform numerical agreement across depths |
| Falcon state-byte check | Depth 1200, single-thread/eager seed 0 agreement reported | Narrow execution setting |
| Concurrent non-preemptive reference repeats | 16 requests: state-byte agreement 6/16; token agreement 11/16 | Reference variability exists in that setting |
| Performance output accounting | Completion and output length checked over 96 pinning cells | Full token sequence not retained for that performance contract |

*Table D-3 (continued)*

| Check | Recorded result or scope | Limit |
|---|---|---|
| Allocator invariants | No reported violations over 24 tested conditions | Not a full retained-output proof |

Values and scope follow the frozen correctness table. Model, protocol, dtype, reference stability, and the object compared differ between rows.

Observable continuation comparisons find a failure that performance results cannot override. In the clean retained-path comparison, both PGK and S2 cases have observable output divergence from the selected-only recomputation reference. The retained dump supports text-SHA comparison, not a silently upgraded direct token-ID test. Equal completion length and success status therefore do not mean equal output content.

Later reference and isolation checks limit attribution rather than erasing that divergence. Five selected-only reference repeats show no observed instability in the tested repeat sets, but the full-workload isolated condition still shows output variability without cross-request retained-state reuse. The corrected targeted-isolation fidelity check has 0/5 eligible PGK pairs and 1/5 S2 pairs, leaving its overall comparison not evaluated. Thus, retained-state corruption is not established and the mechanism remains unresolved. A partial fidelity gate must not be promoted to a directional text verdict.

Code identity and performance accounting also stop short of an execution-wide correctness proof. The handoff reports selected files byte-identical across revisions, while default dispatch behavior can still differ. Similarly, matching completed request counts and output lengths supports performance-volume consistency, not response identity. These limits preserve the meaning of the valid unit tests while keeping the unresolved retained continuation explicit.

# Appendix E. Detailed Results and Sensitivity Analyses

## E.1 Stepwise Performance, Input Accounting, and Slab Storage

The earlier three-seed anchors connect throughput to recorded work without supplying a total-work oracle. Tables E-1 and E-2 preserve their run means and paired relative changes. From S0 to S1, record count K falls more than total recorded tokens H, so H/K rises. That rise does not mean more total work or a measured increase in GPU batch width. These runs are the original inputs to the accounting reanalysis, not extra repetitions added to the twelve-seed extension.

**Table E-1. Earlier high-load anchors: recorded work and throughput**

| Operating point | Org. | Tokens/s | H (tokens/run) | C_once (tokens/run) | K (records/run) | Mean H/K |
|---|---|---|---|---|---|---|
| b63; 3.0 req/s; cap63 | S0 | 283.681 | 839,160.3 | 766,252.3 | 10,500.3 | 79.893 |
| b63; 3.0 req/s; cap63 | S1 | 429.819 | 775,672.3 | 702,764.3 | 4,952.7 | 157.480 |
| b63; 3.0 req/s; cap63 | S2 | 470.308 | 685,389.7 | 612,481.7 | 4,493.0 | 153.895 |
| b108; 4.5 req/s; cap109 | S0 | 417.266 | 770,723.0 | 697,815.0 | 5,469.7 | 141.074 |
| b108; 4.5 req/s; cap109 | S1 | 611.061 | 571,384.3 | 498,476.3 | 2,858.7 | 201.723 |
| b108; 4.5 req/s; cap109 | S2 | 667.210 | 502,947.0 | 430,039.0 | 2,610.0 | 195.272 |

Three seeds. The mean O_ctr is 72,908 tokens/run in these high-load cells, and H=C_once +O_ctr is a counter identity. H/K is the mean of run ratios, not the ratio of displayed means.

**Table E-2. Paired changes in the earlier high-load anchors**

| Operating point | Step | Throughput (%) | C_once (%) | H (%) | K (%) |
|---|---|---|---|---|---|
| b63; 3.0 req/s; cap63 | S0→S1 | +51.4997% | -8.6118% | -7.8233% | -52.8181% |
| b63; 3.0 req/s; cap63 | S1→S2 | +9.4512% | -12.8729% | -11.6423% | -9.3407% |
| b108; 4.5 req/s; cap109 | S0→S1 | +46.6513% | -28.5429% | -25.8427% | -47.7829% |
| b108; 4.5 req/s; cap109 | S1→S2 | +9.2296% | -13.8063% | -12.0146% | -8.7425% |

Each entry averages three same-seed relative changes. Direction agrees in 3/3 seeds for the displayed gain and work reductions, but no new significance test is attached to the range or sign count.

Preemption reduction is not necessary for every observed stepwise gain. At b108,4.5 requests/s, cap 109, seed 1, all three organizations have zero preemptions, while throughput is 439.910,621.652,675.248 tokens/s for S0, S1, S2. This seed is part of the earlier anchor rather than a new independent sample. It excludes preemption reduction as a necessary explanation for that case, but does not make preemption cost zero in other runs.

Low-load accounting does not follow the same input-work direction as the high-load gains. In Table E-3, S1→S2 increases mean paired C_once by 73.5306% at b63 and 99.8321% at b108, while throughput changes only $-0.0086\%$ and $-0.0266\%$. Waiting p50 is zero but transient waiting can still occur, including b63 S0 maxima 5/5/3 across seeds. Accordingly, low-load throughput equality is neither work equality nor an overhead or capacity-equivalence result.

**Table E-3. Low-load input accounting in the original three-seed runs**

| Operating point | Org. | Tokens/s | H (tokens/run) | C_once (tokens/run) | K (records/run) |
|---|---|---|---|---|---|
| b63; cap63 | S0 | 190.020 | 561,997.7 | 489,089.7 | 19,901.7 |
| b63; cap63 | S1 | 190.625 | 259,747.0 | 186,839.0 | 21,258.0 |
| b63; cap63 | S2 | 190.608 | 396,792.3 | 323,884.3 | 20,565.0 |
| b108; cap109 | S0 | 190.634 | 213,667.0 | 140,759.0 | 21,623.3 |
| b108; cap109 | S1 | 190.670 | 186,872.3 | 113,964.3 | 21,629.0 |
| b108; cap109 | S2 | 190.621 | 301,560.3 | 228,652.3 | 21,049.7 |

Arrival rate 0.8 requests/s. These are the original seeds 0–2 and not the extension low-load seeds 42000–42002.

A same-dump convolution calculation supports a limited representation comparison. The source observes 43 live convolution objects in 3 slabs in one original S2 dump: dedicated 512-KiB pages for the same objects would use 21.5 MiB, versus 1.5 MiB in those slabs, a20.0-MiB conditional difference. A separate extension dump has 69 objects in 5 slabs, giving 34.5 versus 2.5 MiB and a32.0-MiB difference. The hypothetical dedicated-page amount is not memory measured in another S1 run. Neither example is a time average or a total hybrid-state maximum.

$$
\begin{aligned}
\widehat{B}_{\text{conv},S1|Z_j} &= 512\ \text{live}_j\ \text{ KiB},\\
\widehat{B}_{\text{conv},S2|Z_j} &= 512\ \text{pages}_j\ \text{ KiB},\\
\Delta\widehat{B}_j &= 512(\text{live}_j - \text{pages}_j)\ \text{KiB}.
\end{aligned}
\tag{E1}
$$

Slab peak values, same-dump counts, and reclaim events carry different information. Peak pages and peak live slots need not occur together and must not be inserted as one state set into the storage equation. In 3,601 original dump lines, the occupancy histogram identities sum pages and occupied slots consistently, but the untimestamped event order cannot create a time-weighted mean or an exact join to polling. The extension reports cross-invalidation and slab pops in 35 of 63 S2 runs, establishing that reclaim coupling occurs without identifying its time cost or lost reuse. Missing slab fields in S0/S1 mean inapplicability, not zero measured activity.

Higher throughput also does not require running-request growth or a falling record count. In the extension b108,3-request/s, cap 63 condition, S1→S2 reduces C_once and waiting while running mean falls and K rises. This differs from the earlier high-load anchor. Combined with the zero-sharing equal-counter case in Section 5.2, it shows why one counter-direction pattern cannot be made a universal mechanism.

**Table E-4. Extension example with higher throughput but lower running mean**

| Quantity | S1 | S2 |
|---|---|---|
| C_once (tokens/run) | 325,016.417 | 166,467.083 |
| K (recorded iterations/run) | 3,931.750 | 4,183.333 |
| Running mean (requests) | 19.327 | 17.442 |
| Waiting mean (requests) | 7.558 | 0.602 |
| Mean run-level E2E p99 (s) | 32.937 | 22.130 |
| Mean throughput (tokens/s) | 650.570 | 708.214 |

b108;3 requests/s;cap 63;q=1;twelve extension seeds. K is a counter record count, not a GPU-forward count.

## E.2 Budget, Load, Submitted Work, and Event Time

The budget results accompany different input-counter patterns at the low and high budgets. At b63 in the extension, D0's C_once is 34,090.667 tokens lower than S2's, so its competitiveness need not be explained as recovering a large work disadvantage through a faster native path. At b108, S2's recorded input is 403,200 tokens lower and its throughput is higher. An earlier nine-seed b108 accounting difference of 98,645 tokens belongs to another load/cap/cohort and is not added to these extension values.

**Table E-5. The original five selected comparisons**

| No. | Model and arrival condition | Cap | Comparator | Seeds | Mean D0 gap (%) | Direction count D0 / comparator | Sequential p | Holm p |
|---|---|---|---|---|---|---|---|---|
| 1 | Falcon b63; Poisson $\lambda$=3.0 | 63 | S2 | 0–5 | +3.64 | 5 / 1 | 0.1250 | 0.3750 |
| 2 | Falcon b108; Poisson $\lambda$=4.5 | 109 | S2 | 0–8 | −5.97 | 0 / 9 | 0.0078 | 0.0391 |
| 3 | Falcon b63; Poisson $\lambda$=1.7 | 8 | S2 | 0–8 | −8.34 | 0 / 9 | 0.0078 | 0.0391 |
| 4 | Falcon b63; ON/OFF mean $\lambda$=1.7 | 63 | S2 | 0–8 | −0.37 | 3 / 6 | 0.5469 | 0.5469 |
| 5 | Granite b160; Poisson $\lambda$=8 | 136 | S0 | 0–8 | −1.09 | 1 / 8 | 0.1484 | 0.3750 |

Mean seedwise 100(D0/comparator −1). The original five-test family, two-look history, and sign-symmetry assumption are preserved, including the Granite comparison.

The submitted-work diagnostic adds same-run event timing without measuring every replayed position or kernel duration. Its joins reproduce 62,221 token-bearing forwards and 10,428,327 valid tokens. Table E-6 reports mean throughput, valid submitted tokens, forward counts, and event elapsed time. Its M/H points use cap 109 and cannot replace the extension's b108 cap 63 result. In the source table, M-D0 event time is 122.314 seconds because displayed run values were averaged first; rounding raw durations after averaging can give 122.315. We preserve the source display rather than relabel this rounding-order difference as a scientific change.

**Table E-6. Same-run work and event elapsed time in the traced diagnostic**

| Point | Org. | Tokens/s | Submitted tokens/run | Token-bearing forwards/ run | Event seconds/run |
|---|---|---|---|---|---|
| L | D0 | 483.282 | 742,691.0 | 4,084.0 | 150.105 |
| L | S2 | 470.212 | 682,188.3 | 4,507.3 | 154.363 |
| M | D0 | 593.082 | 646,808.0 | 2,794.7 | 122.314 |
| M | S2 | 655.681 | 228,556.3 | 4,348.7 | 110.356 |
| H | D0 | 609.826 | 674,925.0 | 2,380.0 | 118.897 |
| H | S2 | 666.414 | 500,940.3 | 2,625.7 | 108.904 |

Three traced seeds per cell. Throughput and work/time ratios are separate estimands; displayed means are not divided to replace the source estimates. Event time is neither kernel-active union nor an independent wall-time component.

Work totals and event elapsed time do not scale in the same proportion. At L, D0 submits about 8.9% more valid tokens while its event sum is about 2.8% lower and its traced throughput is higher. At M, the token ratio is 2.83 but the event-time ratio is about 1.108, while D0 throughput is lower. At H, the token ratio falls to about 1.35 and the performance gap changes again. These are workload-composition and elapsed-time observations, not equal-work native-versus-paged kernel speedups or a unified extension of the old six/nine-seed result.

**Table E-7. Pooled traced work/time ratios and mean paired performance effects**

| Point | Submitted-token D0/ S2 | Event-time D0/S2 | REF gap (%) | TRACE gap (%) | TRACE−REF (pp) |
|---|---|---|---|---|---|
| L | 1.0887 | 0.9724 | +4.129 | +2.955 | −1.175 |
| M | 2.8300 | 1.1084 | −8.935 | −9.519 | −0.584 |
| H | 1.3473 | 1.0918 | −7.451 | −8.341 | −0.890 |

Work and event ratios pool the same TRACE seeds. Throughput gaps average seedwise relative ratios. Small differences between rounded display subtraction and the reported gap change reflect source precision.

Phase differences describe different execution compositions rather than isolated costs. At M, D0 has a larger mixed-phase event sum and a smaller decode-only sum, but call counts and work shapes also differ. Mixed phases include input work, so a small prefill-only sum does not mean all prefill cost is small. Corrected chunk prefill/decode-input labels come from the pre-update schedule, while planned prompt and generated-history replay are separate quantities and are not extra measured tokens to add to V_sub.

**Table E-8. D0 minus S2 CUDA-event elapsed sums by phase**

| Phase | L seconds/run | M seconds/run | H seconds/run |
|---|---|---|---|
| Prefill-only | +0.0003 | −0.0113 | −0.0043 |
| Decode-only | −7.8407 | −27.0173 | −4.9233 |
| Mixed | +3.5827 | +38.9873 | +14.9197 |
| ALL | −4.2573 | +11.9580 | +9.9927 |

Three common TRACE seeds. Phase sums are elapsed intervals for scheduled work, not kernel-specific times or independent additive wall-time components.

**Table E-9. Chunk work labels and separately planned resume quantities**

| Point | Org. | Chunk prefill tok/run | Chunk decode-input tok/run | Planned prompt replay tok/run | Planned generated-history replay tok/run |
|---|---|---|---|---|---|
| L | D0 | 670,107.7 | 72,583.3 | 41,848.0 | 315.7 |
| L | S2 | 609,580.3 | 72,608.0 | 0.0 | 0.0 |
| M | D0 | 574,214.3 | 72,593.7 | 30,291.0 | 125.0 |
| M | S2 | 155,948.3 | 72,608.0 | 0.0 | 0.0 |
| H | D0 | 602,338.3 | 72,586.7 | 38,167.7 | 139.7 |
| H | S2 | 428,332.3 | 72,608.0 | 0.0 | 0.0 |

Three-seed means. Planned replay is not independently measured completed replay and is not added again to submitted work. Zero planned replay in these S2 runs does not imply zero replay in every campaign.

Most recorded D0 allocation failures involve insufficient positive attention free space rather than a literally empty pool. Table E-10 distinguishes need>free>0 failures, free=0 failures, and successes with no added need. Recurrent shortages occur only sparsely. Counts refer to calls reaching the recorded check and may repeat the same request; outcome labels do not independently identify the exact preemption victim or a complete request-level causal path.

**Table E-10. D0 recorded allocation conditions and following outcome labels**

| Point | A fail:need>free>0 | A fail:free=0 | Success:A free=0 | R fail:need>free>0 | Defer label | Preempt label |
|---|---|---|---|---|---|---|
| L | 3,024.0 | 47.3 | 369.0 | 0.3 | 3,044.7 | 26.7 |
| M | 974.7 | 21.0 | 212.3 | 0.3 | 976.3 | 19.3 |
| H | 1,384.0 | 31.7 | 327.3 | 0.3 | 1,391.7 | 24.3 |

Mean events/run over three TRACE seeds, not failure probabilities over all allocator calls or unique-request counts.

Byte scope must be normalized before interpreting the diagnostic's pool values. The recurrent emitter weights 44 layers, while the comparison planner value is per-layer. Thus 1,594,368 bytes/layer becomes 70,152,192 bytes per logical block over 44 layers; the example 561,217,536 bytes corresponds to eight such blocks. Comparing that weighted value directly with a per-layer plan would create a false inconsistency. Even after normalization, the present records leave completed replay time, kernel-active union, and equal-work native/paged cost unidentified.

## E.3 Auxiliary Concurrency-Cap Comparisons

The cap sweeps show different non-monotonic organization responses at the two budgets. At b63 and 1.7 requests/s, the three-seed cap 3/5/6/8 curve does not change monotonically in relative gap. Comparing cap 8 with the reused cap 63 reference over six common seeds widens S2's advantage by 7.74 percentage points, rather than reversing the ranking. At b108, the gap narrows as the cap relaxes from 5 to 13. The final nine-seed cap 8 test remains distinct from both the three-seed curve and six-seed contrast.

**Table E-11. Cap response and D0 internal-check events**

| Budget | Cap | Source | Mean D0/S2 gap (%) | Rejection / completion | Victim / completion | Preemption / completion |
|---|---|---|---|---|---|---|
| b63 | 63 | Reused TPS2-L reference | −0.98 | 0.199 | 0.032 | 0.032 |
| b63 | 3 | TPS2-C | −3.48 | 0 | 0 | 0 |
| b63 | 5 | TPS2-C | −5.88 | 0 | 0 | 0 |

*Table E-11 (continued)*

| Budget | Cap | Source | Mean D0/S2 gap (%) | Rejection / completion | Victim / completion | Preemption / completion |
|---|---|---|---|---|---|---|
| b63 | 6 | TPS2-C | −9.06 | 0 | 0 | 0 |
| b63 | 8 | TPS2-C | −7.00 | 0 | 0 | 0 |
| b108 | 109 | Reused TPS2-L reference | −0.43 | 0 | 0 | 0 |
| b108 | 5 | b108 concurrency extension | −8.59 | 0 | 0 | 0 |
| b108 | 8 | b108 concurrency extension | −7.82 | 0 | 0 | 0 |
| b108 | 10 | b108 concurrency extension | −4.22 | 0 | 0 | 0 |
| b108 | 13 | b108 concurrency extension | −2.66 | 0 | 0 | 0 |

Arrival 1.7 requests/s;seeds 0–2. References are actual 63/109-cap runs, not unlimited runs. Cap 8 is −7.00% here, −7.86% for six common seeds, and −8.34% in the nine-seed test.

Zero recorded pool rejections inside the swept caps does not mean that checking is disabled or free. The b63 cap 63 reference has attention shortage and rejection events, while the smaller swept conditions record none. Within the tested b63 range, higher caps reduce end-to-end p99, showing the cost of queueing under a small cap without validating arbitrary larger caps. Request selection, queueing, and state demand change together, so the sweep does not isolate pool separation or one resident-request count.

## E.4 Arrival Bursts and Tail Latency

Arrival concentration changes queueing even when mean load and cap are held fixed. At b63, mean 1.7 requests/s, and cap 63, the six-seed mean of run-specific waiting-count p99 rises from 8.0 to 22.1 for D0 and 6.4 to 18.3 for S2 when Poisson arrivals become five-second-on/fifteen-second-off arrivals. These are counts from polling, not a request queue-delay p99.

Inconclusive throughput under burst arrivals does not establish latency equivalence. The nine-seed within-burst D0/S2 gap is −0.37% with adjusted p=0.5469, while E2E p99 and request-TPOT p99 ratios are 1.171 and 1.857. Those are within-pattern ratios, not the increase caused by switching from Poisson to burst arrivals. The six-seed pattern comparison and nine-seed organization test are kept separate, and no new tail-latency significance test is assigned.

## E.5 D0 Pool-Split Sensitivity

The split sweep tests whether the calibrated default also happens to maximize measured serving throughput. At 3 requests/s, cap 63, q=1, and twelve seeds, the attention fraction takes five values around r0=0.527383 at b63 and b108. All eight non-default-versus-default contrasts reject in the original 22-test family. The highest observed mean is at r=0.477383 for both budgets, so the source does not support an optimum shifting between these two budgets.

**Table E-12. D0 non-default splits relative to the calibrated default**

| Budget | Setting | Attention fraction | Mean relative effect (%) | Holm p |
|---|---|---|---|---|
| b63 | RM10 | 0.427383 | -2.480 | 0.01074 |
| b63 | RM05 | 0.477383 | +0.850 | 0.03955 |
| b63 | RP05 | 0.577383 | -4.311 | 0.01074 |
| b63 | RP10 | 0.627383 | -10.959 | 0.01074 |

*Table E-12 (continued)*

| **Budget** | **Setting** | **Attention fraction** | **Mean relative effect (%)** | **Holm p** |
|---|---|---|---|---|
| b108 | RM10 | 0.427383 | +0.867 | 0.03662 |
| b108 | RM05 | 0.477383 | +1.854 | 0.01074 |
| b108 | RP05 | 0.577383 | -6.150 | 0.01074 |
| b108 | RP10 | 0.627383 | -13.779 | 0.01074 |

Original 22-test family; twelve paired extension seeds. RM05 is the observed best of the tested values, not a global optimum or an independent held-out selection.

The split changes rounded byte totals as well as where they reside. At RM05, the planner total rises 1.320% at b63 and falls.402% at b108 relative to the default. Subtracting those byte percentages from throughput percentages would not isolate a pure split effect. Increasing the attention fraction toward RP10 instead shifts most recorded shortage events to the recurrent-only category: pooled shares are 87.934% and 86.458% at the two budgets, while means of run-specific shares differ. These are compositions of recorded failure events, not overall failure probabilities.

The observed best split does not remove the b108 S2 advantage in the supplied comparison. D0/S2 changes from +4.978% to +5.867% at b63 and from $-12.556\%$ to $-10.941\%$ at b108 when the observed RM05 setting is used. Because RM05 was selected from these same measurements, no new superiority test is attached to this comparison. The result supports only the tested split sensitivity, not a globally optimized D0 baseline.

## E.6 Prefix-Sharing Interactions

The sharing-endpoint result does not describe the whole sharing curve. The D0/S2 endpoint interaction of +1.566 percentage points is inconclusive, yet at sharing fraction.5 the mean D0/S2 effect is +3.291% with 12/12 positive seeds and an observed range +0.893% to +5.149%. These intermediate observations remain descriptive and are not added after the fact to the primary family or interpolated into a continuous crossover point.

Candidate reuse and actual submitted work still differ in this sharing analysis. A first-lookup candidate depth, a once-recorded input counter, and GPU work count different things. The sharing results can be connected to performance at their recorded scope, but candidate-hit changes alone do not establish successful adoption, eviction causality, or a time contribution. Table G-2 retains the three sharing levels and the original endpoint tests.

## E.7 Auxiliary Granite Model Scope

The auxiliary Granite comparison is not a second-model replication of the Falcon stepwise experiment. Its later two-arm study compares D0 with S0, the most competitive shared design in its earlier measurements. At b160 and 8 requests/s, the D0/S0 mean gap is $-1.09\%$ and is not established after correction. The earlier three-seed four-arm Granite experiment reports S1/S2 below S0 under its own protocol. We preserve that result as a generalization limit rather than treating the presence of a second model as universal support for the Falcon ranking.

## E.8 Latency-Qualified Goodput Sensitivity

Adding empirical latency thresholds can change the apparent service-performance gap. The auxiliary analysis uses thresholds derived from either the comparator requests or their union, then applies them to the same observed requests. TableE-13 shows that burst and Granite conditions can retain a latency-qualified disadvantage for D0 even where plain throughput is inconclusive. These thresholds are not independently chosen application SLOs.

**Table E-13. Latency ratios and empirical SLO-goodput sensitivity**

| Model / comparator condition | Seeds | E2E p99 ratio | TPOT p99 ratio | Comparator-threshold effect (%) | Union-threshold effect (%) |
|---|---|---|---|---|---|
| Falcon b63; $\lambda$=3.0; cap63 / S2 | 0–5 | 0.944 | 1.155 | +3.57 | +5.08 |
| Falcon b108; $\lambda$=4.5; cap109 / S2 | 0–8 | 1.159 | 1.367 | −27.63 | −17.43 |
| Falcon b63; $\lambda$=1.7; cap8 / S2 | 0–8 | 1.464 | 1.167 | −58.93 | −29.33 |
| Falcon b63; ON/OFF mean $\lambda$=1.7; cap63 / S2 | 0–8 | 1.171 | 1.857 | −17.45 | −8.48 |
| Granite b160; $\lambda$=8; cap136 / S0 | 0–8 | 1.023 | 1.243 | −6.92 | −5.27 |

Thresholds are estimated from the evaluated requests themselves. These are post hoc service-metric sensitivities, not new independently preregistered SLO guarantees.

The size of a latency-qualified difference depends on the selected threshold. For b108 high load and b63 cap 8, the comparator-threshold gaps are larger than plain throughput gaps. This does not decompose the original gain into a latency mechanism. The comparator and union views are retained together to show that dependence rather than choosing the more favorable result.

## E.9 Full-Run and Central-Arrival Views

The central-arrival analysis is not a steady-state capacity estimate. Of 84 runs in the original five comparisons, 83 fail the central-window drift heuristic. Mean gap directions survive in several comparisons but their sizes change, and at b108 one of nine seed directions changes. Burst arrivals even change the mean gap sign in the central view. Consequently, neither full-run p-values nor direction consistency is transferred to the central window.

**Table E-14. Mean performance gaps over different observation windows**

| Comparison | Seeds | Full-run gap (%) | Central-arrival gap (%) |
|---|---|---|---|
| Falcon b63; $\lambda$=3.0; cap63; D0–S2 | 0–5 | +3.64 | +4.83 |
| Falcon b108; $\lambda$=4.5; cap109; D0–S2 | 0–8 | −5.97 | −13.90 |
| Falcon b63; $\lambda$=1.7; cap8; D0–S2 | 0–8 | −8.34 | −9.07 |
| Falcon b63; ON/OFF mean $\lambda$=1.7; cap63; D0–S2 | 0–8 | −0.37 | +0.57 |
| Granite b160; $\lambda$=8; cap136; D0–S0 | 0–8 | −1.09 | −0.80 |

The central view is a separate window on the finite trace, not an independent sample or steady-state test. Full-run significance is not reassigned to these values.

The submitted-work diagnostic uses a different central-window reconstruction. It derives a20th–80th percentile arrival window within each realized TRACE run, rather than using the original common frozen-arrival window. At M, D0 submits more tokens while the event sums in those fixed-length windows are similar. Similar interval totals do not imply equal whole-request latency or equal-work kernel speed. The diagnostic's central values therefore remain subordinate to its full traced record.

## E.10 Connecting Work, Reuse, and Time without Merging Their Units

The separate work records jointly rule out a single universal input-reduction explanation. The earlier high-load input counter falls with better throughput; the low-load counter can rise while throughput stays similar; and zero-sharing extension cells have identical recorded input but different throughput. Submitted-work

diagnostics then add a different recording level. These results should be connected as condition-specific observations rather than pooled into one workload or one total-work metric.

No single scheduler or allocation counter substitutes for effective capacity or kernel efficiency. Record count K is not a GPU-forward count, H/K is not compute efficiency, and sampled running requests are not maximum state capacity. The same-dump slab calculation supports a limited representation comparison, but lacks the complete simultaneous attention and allocator state needed for a measured capacity maximum. Split and need/free diagnostics also have different cohorts and scopes.

The evidence therefore connects storage, reuse, work composition, and elapsed time while leaving causal fractions unresolved. A higher submitted-token count need not produce a proportional event-time increase. Phase timing does not reveal completed replay's time share or equal-tensor native/paged cost. The defensible conclusion is the observed relation within each compatible experiment, not a fixed percentage of performance assigned to one cause.

## E.11 Instrumentation Effects

The original logging check finds small sample means at one operating point rather than a bound on all measurement overhead. At b108,4.5 requests/s, cap 109, the twelve-run three-seed comparison reports throughput changes of +0.30% for D0 and +0.31% for S2, and step-time changes of −0.12% and −0.20%. Both modes retain one-Hz polling, and order rotates by seed. Poller cost and effects outside this point are therefore not established as zero.

The separate work-timing probe has variable effects that a signed mean can hide. Across eighteen reference/traced pairs, the mean throughput change is +0.3584% but individual values range from −1.3115% to +5.4587%. D0–S2 mean gaps change by −1.175, −0.584, −0.890 percentage points at L, M, H. In the L-S2-seed 1 example, TTFT p50 moves from 13.252 to 6.181 seconds and E2E p50 from 24.587 to 17.811 seconds. These observed changes are not all causally assigned to instrumentation, but they prevent a claim that every latency effect is small.

Separate profiler captures are preparation for a future fixed-work comparison, not a full serving-time decomposition. The microcapture and descriptor candidates are excluded from the 36 live diagnostic runs. Their operation fractions cannot be used as the complete native-path advantage or a bound on serving speedup.

## E.12 Stranded Bytes and Redistribution Screening

The original stranded-byte screen observes little step-weighted idle space in the tested D0 configurations. Across 117 runs, the maximum is.09% for Falcon load,0.06% for bursts, and.19% overall in Granite. The metric measures free bytes in one pool when the other is exhausted, not total utilization, instantaneous rejection probability, or proof of no memory bottleneck. The separate extension reports maxima.37% in final cumulative dumps and.66% during cumulative progress;those scopes are not merged.

The decision not to build a dynamic redistribution prototype follows an engineering screen, not a measured performance theorem. The observed maximum is below the.40% lower end of a prior host-side virtual-memory/remapping break-even model. That threshold is not a preregistered measured-throughput boundary. The screen therefore does not reject dynamic redistribution for other splits, workloads, or implementations.

## E.13 Preemption-Time Checkpoint-Pinning Sensitivity

The 96-cell pinning sweep asks whether the tested layout gap depends strongly on one preemption-time setting. It combines the two operating points, two organizations, four rules, and six paired seeds described in B.2. D0 pin counts change from zero under Roff to roughly 26–28 per run under enabled rules. S2 has few preemption or pin opportunities, with small nonzero activity in some seeds. Cumulative events are not occupancy, and the same logical threshold need not produce the same realized intervention.

**Table E-15. Mean D0–S2 gaps across preemption-pinning settings**

| Operating point | Cap | Roff:pin off | R0:K3 | R1:K2 | R2:K1 |
|---|---|---|---|---|---|
| A: Falcon b108; $\lambda$=4.5 | 109 | −4.74% | −4.46% | −4.74% | −4.87% |
| B: Falcon b63; $\lambda$=3.0 | 63 | +3.62% | +3.24% | +3.64% | +3.68% |

Six paired seeds per setting. Ordinary recurrent prefix caching remains enabled when extra preemption-time pinning is off. R1 uses 24 compatible older cells;72 other cells are fresh.

**Table E-16. Primary pinning interaction and log-ratio sensitivity**

| Operating point | Contrast | Mean change (pp) | Observed seed min/ max (pp) | Raw relative-gap p | Raw log-ratio p |
|---|---|---|---|---|---|
| A: b108; $\lambda$=4.5; cap109 | R0→R2 | −0.4097 | −1.8029 / +1.3317 | 0.43750 | 0.46875 |
| B: b63; $\lambda$=3.0; cap63 | R0→R2 | +0.4425 | −1.0253 / +2.2184 | 0.43750 | 0.43750 |

The two primary raw p=0.43750 values become Holm p=0.875 in the separate two-test family. The log-ratio column is sensitivity, not a replacement primary.

Small average pinning sensitivity does not establish equivalence. The primary R0→R2 mean changes are −0.4097 percentage points at A and +0.4425 at B, with adjusted p=0.875 for both. All eight point–contrast mean magnitudes are below.5 points, but the largest individual-seed magnitude is 2.4707 points. The.5-point observation is therefore not an equivalence bound. The result applies to this preemption-pinning ladder, not checkpoint placement, producer-running retention, all eviction policies, or other models and pressures.

## E.14 Run Adoption, Statistical Sensitivity, and Reproduction Limits

The extension's adopted rerun remains a documented protocol deviation. The final analysis has 348 unique completed runs, but order139—S1, b108, q1,3 requests/s, seed 42004—first completes 299/300 and is rerun alone about 15 hours later. Some first-attempt raw is overwritten and the failure cause is unresolved. Adopting the later complete run does not certify execution of the originally intended block-level retry plan.

**Table E-17. Adoption scenarios retained for the extension family**

| Scenario | Treatment of affected contrasts | Rejections in family 22 | Interpretation |
|---|---|---|---|
| A:current adopted analysis | Use documented attempt 2; twelve seeds per main contrast | 20/22 | Main reported effects and p-values |
| B:complete-case sensitivity | Exclude 42004 from the two affected contrasts; n=11 there | 20/22 | Post hoc support, not replacement of A |
| C:invalid-adoption assumption | Affected effects not estimable; assign p=1 for correction | 16/22 | Conditional scenario, not an actual invalidity decision |

The rerun affects S1/S0 and S2/S1 at b108 and can affect Holm adjustment indirectly. Table B-6 reports the available n=12/n=11 values.

The freeze history also includes a pre-final-freeze observation and cannot be described as entirely outcome-blind. A 300-request driver smoke used a formal seed before the final freeze, and its original data is lost. Earlier design and contrast records remain evidence of planning, but do not erase that observation. The main adoption, seed-exclusion sensitivity, and invalid-adoption scenario are all retained rather than choosing only the favorable one.

Later artifact delivery strengthens particular checks without restoring missing historical executions. The later handoff links source effects and adjusted tests, and B.12 reports their consistency. It does not recover the original failed request/poller/server record or every missing historical model and binary identifier. Earlier source statements about unavailable files are historical statuses; the later F.5 scope controls what was

actually supplied. Raw-to-table reproduction for one accounting or timing campaign is not automatically transferred to the extension.

The remaining unidentifiable quantities are not filled by local validation success. The input-counter overwrite classification has a conditional bound of about 0.57% under its membership rule, not a bound on true GPU work, throughput, or model error. The timing analyzer passes the supplied 19 and separate 12 regression scenarios and regenerates 13 CSVs, but this is not proof for every input. Completed replay volume and time, kernel-active union, and equal-state native/paged cost remain unidentified. The manuscript introduces no new GPU run or fixed-work probe to fill those gaps.

## Appendix F. State-History Analysis and Evidence Provenance

### F.1 From Existing Metrics to a State Trace

The trace work begins with a limitation of the existing performance metrics: they cannot distinguish preserved state from state newly available during waiting. The initial offline review uses the 360-run admission-confirmation record and 108,000 requests without adding GPU runs. Candidate cache-hit fields do not establish actual adoption or connect eviction with a later request's lost reuse. Instrumentation is then added to record the waiting-entry hash chain and publication, loss, adoption, and allocation-cause context.

The precursor validates the observation method before the final headroom comparison. An S2 functionality study uses 120 requests under performance, trace, and repeated-performance modes with one seed and retention off. Its trace reconstructs 120/120 entry and adopted depths, with 66 zero-to-zero, 25 positive-equal, 16 newly reusable, one deeper, and 12 reduced-depth episodes. A displayed $-0.39\%$ throughput change at one seed cannot pass the later three-seed fidelity rule. D0's 16/1792 geometry has separate synthetic fixtures rather than inheriting an S2-only replay assumption.

The final trace closes the feasible observation chain while retaining its structural gaps. The formal 24-run comparison uses native D0 and S2, common headroom settings and three seeds, with paired performance and traced modes. Collection and fidelity pass, and the corrected endpoint, entry-reconstruction, scalar replay, and lineage results are those in Appendix C.1. Producer identity and full lookup-map snapshot equivalence are not established. The source therefore closes its current descriptive analysis but retains the broader mechanism question as partial, not as a failed collection or a complete causal proof.

### F.2 Evidence Sources and the Role of Their Records

Evidence records are grouped by what they establish rather than by an assumed universal pass status. Configuration and runner audits identify layout, admission, and budget rules. Workload manifests and adopted ledgers identify executed cells and their exceptions. Request, poller, slab, and trace records support particular result reconstructions. Corrections and final gates define which earlier records are superseded. The source map accompanying this draft connects manuscript paragraphs and tables to the corresponding frozen anchors.

Historical diagnostics remain separate from the current native D0–S2 comparison. The boundary-matched input-work, isolation, scheduling, time-instrumentation, fixed-work, and live separation studies often use D0-M512. They supply information about different controls and measurement levels, but their absolute performance and timing must not be imported into the native comparison. Similarly, the original headroom development, independent confirmation, and later state trace retain their own seeds and protection settings.

Retry and correction records are part of the evidence rather than optional editorial detail. The admission confirmation introduces a bounded connection-error retry amendment during collection, and the adopted data includes one request-internal retry whose original server-side duplicate or cancellation behavior is not recorded. The source retains the n=12 primary and separate event-seed sensitivity. State-history replay also required corrected request-plus-preemption binding; the superseded request-only lineage is not reused as

the final count.

### F.3 Current Results and Superseded Reports

The current state-history result is the corrected episode-bound analysis, not every earlier number in its source history. Earlier 357-episode shrink counts, incomplete replay matches, and request-ID-only overwrites are historical stages. The current block-ID reconstruction has 385 reduced-depth episodes, and the corrected scalar endpoint and lineage counts match 3,714 episodes. Figure 7(c) and Table G-7 report Q9 retention performance, not Q7 event-lineage evidence. The source map preserves separate endpoint evidence without changing old files in place.

Descriptive closure does not identify every producer or establish a unique performance cause. The corrected trace establishes inclusion of other-request loss in each reduced-depth episode at the defined scalar-depth level. It does not reconstruct the entire cache at every lookup or attribute all throughput differences to that loss. Older progress labels are therefore retained only in the authoring archive and are not repeated as current manuscript conclusions.

### F.4 Manuscript, Reproduction Artifacts, and Authoring Records

A checked evidence handoff is not itself proof of scientific correctness or public reproducibility. The source reports matching the size and SHA256 of 90 designated files, with 91 ZIP members including the manifest. Those checks establish the delivered copies. They do not supply all original logs, the complete historical engine repository, every model identity, or an independent execution replication.

The manuscript and the authoring evidence retain different material. Scholarly text, necessary appendices, and public references belong in the paper. Run scripts, frozen manifests, analyzers, statistical inputs, and representative logs that were supplied are no longer marked missing merely because an earlier version lacked them. Full run directories, 39 frozen input traces, and complete K3 JSON remain outside the supplied scope. Internal paragraph IDs, review questions, intermediate classifications, and historical tables are kept in the authoring evidence rather than silently presented as extra current experiments.

This draft preserves the frozen source separately from its manuscript rendering. Each manuscript paragraph and table has a source-map entry, and the complete source-table archive retains original text and hashes for material not re-typeset. This editorial separation follows the source's distinction between necessary paper tables and the complete working ledger. It does not decide public release of private repositories or claim that the draft bundle alone can rerun every experiment.

### F.5 Delivered Evidence, Corrections, and Remaining Gaps

The later handoff resolves specified evidence links without retroactively certifying unrecorded fields. It supplies execution and analysis artifacts, representative startup indexes, statistical effect inputs, and model/workload scope records. This supports the kernel-path and statistics connections described in Appendix B. The historical contract typo is handled through a hashed derivative, and the six representative sharing-level logs are connected without calling that a new review of every boot record.

Missing full raw and identity fields remain explicit even after the local document is ready for drafting. Runtime usable-memory snapshots, historical model hashes, all frozen input files, and original failed-attempt raw do not become known by copying a later environment record. Conversely, already-delivered analyzers and statistical inputs should not keep an obsolete missing-data label. The frozen evidence supports a scope-specific result and handoff account, not a blanket claim that every historical execution or public reproduction dependency is complete.

# Appendix G. Supplemental Tables and Figure-Reading Notes

## G.1 Recorded Work, Displayed Indices, and Sharing Levels

This appendix restores result tables needed to read the revised Figures 3–7. It uses the supplied long-manuscript tables and Q8/Q9 audit outputs; no new serving runs or statistical families are introduced. The compact Fig. 3(b) preserves the mean within-seed S0-normalized throughput index, while Fig. 7(c) retains all Q9 seed observations; source measurements and statistical estimands are unchanged.

**Table G-1:** Q2 no-shared-context comparison at b83

| Org. | Initial-input tokens/run | Recorded total tokens/run | Preemptions per run | Output tokens/s (arithmetic mean) |
|---|---:|---:|---:|---:|
| S0 | 1,009,603.083 | 1,088,065.667 | 0 | 293.836 |
| S1 | 1,009,603.083 | 1,088,065.667 | 0 | 405.376 |
| S2 | 1,009,603.083 | 1,088,065.667 | 0 | 415.862 |

Exp. 2c: $q = 0$, 3 req/s, cap 63, twelve seeds. Underlying values are reproduced in Table G-1. Equal recorded counters do not establish equality of all GPU operations.

**How to read Figure 3(b).** Let $e^{(1)}_{1,s}$ and $e^{(1)}_{2,s}$ be the archived q=1 seedwise relative effects (percent) for S1/S0 and S2/S1, and let $i_{1,s}$ and $i_{2,s}$ denote the corresponding q=1-minus-q=0 interaction effects (percentage points). For the same twelve paired seeds $s = 42000, \ldots, 42011$, recover the q=0 step effects as $e^{(0)}_{j,s} = e^{(1)}_{j,s} - i_{j,s}$. For $r_{j,s} = 1 + e^{(0)}_{j,s}/100$, the S0-normalized indices are

$$
\begin{aligned}
I_{\mathrm{S0}} &= 1, \\
I_{\mathrm{S1}} &= \frac{1}{12}\sum_s r_{1,s} = 1.379451456, \\
I_{\mathrm{S2}} &= \frac{1}{12}\sum_s r_{1,s} r_{2,s} = 1.415129435.
\end{aligned}
$$

Crucially, the S2/S0 product is taken *within* each matched seed before averaging. This is not the product of separately averaged step ratios (1.415119728 with unrounded step means), nor a ratio of the three absolute run-mean throughputs in Table G-1. The archived source tables are `RECOMPUTED_CONTRAST_SEEDS.csv` (twelve complete seedwise rows for each of the four contrasts) and `PREFIX_DESCRIPTIVE_ALL_Q.csv`; the computed per-seed ratios are included as `Q2_Q0_MATCHED_SEED_RATIOS.csv` in the figure-freeze package. The corresponding indices round to 1.000/1.379/1.415; numeric marker annotations are omitted from the compact Figure 3(b). No absolute per-run throughput is reverse-engineered and no kernel-level cause is inferred from these contrasts.

**Table G-2:** Q3 relative throughput at all three shared-context fractions

| $q$ | Mean S1/S0$-1$ (%) | Mean S2/S1$-1$ (%) | Mean D0/S2$-1$ (%) |
|---|---:|---:|---:|
| 0 | +37.945 | +2.586 | $-3.294$ |
| 0.5 | +38.797 | +4.150 | +3.291 |
| 1 | +47.568 | +12.628 | $-1.728$ |

Exp. 3a: b83, 3 req/s, cap 63, twelve paired seeds. Underlying values are reproduced in Table G-2. Entries are means of seedwise relative effects, not interpolated values. Reused runs are not additional samples.

Figure 4 intentionally shows only the two endpoints. Its $q = 0$ S2/S1 label is corrected to 2.586%, the three-decimal rounding of the source mean 2.5856851625%. The 10.043-percentage-point endpoint interaction remains a separately computed paired statistic rather than subtraction of independently rounded bar labels. The reported endpoint interactions remain +9.623 and +10.043 percentage points, both with Holm-adjusted $p = 0.01074$. The D0/S2 endpoint interaction remains +1.566 points ($p = 0.41113$); the middle point is descriptive, not a new crossover or superiority test (Appendix E.6).

## G.2 Q7 Precision and Q8 Absolute Depths with Reconstruction Brackets

In Figures 7(a,b), each depth is a mean over requests within the stated arm–seed scope, not the largest request depth in the run. $A_{\text{max}}$ is the deepest recurrent-checkpoint boundary supported by the request's contiguous attention prefix, $R_{\text{max}}$ is the deepest complete recurrent checkpoint, and $J_{\text{adopt}}$ is the scheduler-adopted joint depth. Lookup-time supply is bracketed by M0 and M1; the plotted $A_{\text{max}}$ and $R_{\text{max}}$ use M1. A reconstruction bracket is not a statistical confidence interval.

**Q7 source precision.** Figure 7(a) shows separately rounded S2 $J_{\text{adopt}} = R_{\text{max}}$ means over seeds 76600, 76601, and 76602, respectively. For $f = 0$ they are 1338.0, 1583.8, and 1705.0; for $f = 0.19$ they are 2123.1, 2234.0, and 2259.6 tokens/request. The seedwise paired increases are 785.1, 650.2, and 554.7 tokens/request. The 0.1-token difference for seed 76602 comes from subtracting the *displayed* means rather than the underlying quantized state depths. Under the reported 512-token boundary grid and 300 INITIAL requests per arm, the unique grid-consistent sums imply 999 and 1324 units. Thus $(1324-999)\times512/300 = 554.666667$ tokens/request and rounds to **554.7**, whereas $2259.6-1705.0 = 554.6$ subtracts two previously rounded values. This is a static grid-and-denominator reconstruction, not a replay of per-request raw event histories.

**Table G-3:** Q8 arm–seed mean absolute depths, in tokens/request

| Seed | Arm | $J_{\text{adopt}}$ | $R^{\text{M0}}_{\text{max}}$ | $R^{\text{M1}}_{\text{max}}$ | $A^{\text{M0}}_{\text{max}}$ | $A^{\text{M1}}_{\text{max}}$ |
|---|---|---|---|---|---|---|
| 79010 | S2 | 2,266.45 | 2,266.45 | 2,266.45 | 2,918.40 | 2,918.40 |
| 79010 | S2+Ret. | 2,969.60 | 2,969.60 | 2,969.60 | 2,969.60 | 2,969.60 |
| 79011 | S2 | 1,472.85 | 1,472.85 | 1,472.85 | 2,809.17 | 2,816.00 |
| 79011 | S2+Ret. | 2,911.57 | 2,911.57 | 2,916.69 | 2,928.64 | 2,938.88 |
| 79012 | S2 | 1,846.61 | 1,846.61 | 1,846.61 | 2,856.96 | 2,867.20 |
| 79012 | S2+Ret. | 2,938.88 | 2,938.88 | 2,938.88 | 2,938.88 | 2,938.88 |

S2+Ret. denotes S2+Retention. Both arms use 254 pages, SHARE, 3 req/s, cap 63, and $f = 0$. Each cell averages 300 workload-index-paired INITIAL requests. The 25 RESUME episodes in the seed-79011 retention arm are excluded under the original analysis rule. Values are displayed to two decimal places; calculations use the supplied higher-precision values.

**Table G-4:** Q8 paired depth increases and attention-availability brackets

| Seed | $\Delta J_{\text{adopt}}$ | $\Delta R^{\text{M1}}_{\text{max}}$ | $\Delta A^{\text{M1}}_{\text{max}}$ | $\Delta A_{\text{max}}$ bracket |
|---|---|---|---|---|
| 79010 | +703.15 | +703.15 | +51.20 | $[+51.20, +51.20]$ |
| 79011 | +1,438.72 | +1,443.84 | +122.88 | $[+112.64, +129.71]$ |
| 79012 | +1,092.27 | +1,092.27 | +71.68 | $[+71.68, +81.92]$ |

S2+Retention minus S2, tokens/request. The attention bracket is $[A^{\text{M0}}_{\text{Ret}} - A^{\text{M1}}_{\text{S2}},\ A^{\text{M1}}_{\text{Ret}} - A^{\text{M0}}_{\text{S2}}]$, not an effect CI.

**The seed-79011 asterisk.** In the retention arm, $R^{\text{M1}}_{\text{max}} = 2916.693333$ and $R^{\text{M0}}_{\text{max}} = J_{\text{adopt}} = 2911.573333$. The 5.12-token mean gap comes from one bounded request with $1536/300 = 5.12$. The reported $\Delta\text{Gap}_R \in [0, 5.12]$ is not an exact loss. The other five cells have equal $J_{\text{adopt}}$ and $R_{\text{max}}$ means. Figure 7(b) uses small horizontal offsets to separate coincident metrics, not different observation times.

Attention availability also increases in every seed, with positive lower brackets; it is not held invariant by the intervention. Its gains are smaller than the recurrent/joint gains. These observations do not identify a mediation fraction or establish retained-state continuation correctness.

Sources: supplied `Q8_ABSOLUTE_JR_VALUES.csv`, `Q8_AMAX_BY_SEED.csv`, and their audit reports. Current checks reproduce arithmetic relationships among these extracts; raw event reconstruction was reported by the experiment session, not rerun in this editorial revision.

### G.3 Q9 Absolute Ledger Throughput and Complete Seed Values

Q9 is the separate fresh-seed S2/S2+Retention $\times$ ISOLATE/SHARE confirmation at 254 pages, 3 req/s, cap 63, and $f = 0$. The primary ledger throughput is completed successful output tokens divided by the runner's ledger episode duration. This wall-clock interval includes benchmark orchestration and completion polling, while excluding server boot and shutdown; it is not pure GPU time. Request-span throughput remains a separate sensitivity analysis.

**Table G-5:** Q9 per-seed ledger throughput: all 24 cells

| | ISOLATE | | SHARE | |
|---|---|---|---|---|
| Seed | S2 | S2+Retention | S2 | S2+Retention |
| 88001 | 367.534 | 369.327 | 522.152 | 560.830 |
| 88002 | 381.056 | 381.056 | 498.836 | 715.722 |
| 88003 | 374.522 | 374.522 | 529.497 | 568.719 |
| 88004 | 384.449 | 384.449 | 477.648 | 583.793 |
| 88005 | 376.463 | 374.636 | 571.667 | 571.667 |
| 88006 | 345.845 | 363.490 | 487.973 | 565.429 |

Output tokens/s, one run per arm–seed, six paired seeds. Display precision is three decimals; summary calculations retain the supplied full precision. No row is removed because two policies have the same recorded throughput.

**Seed visibility in the confirmed Figure 7(c).** The SHARE/S2+Retention observation at seed 88002 is 715.721739 output tokens/s. The marker is drawn at the measured coordinate above the legend layer in Fig. 7(c). The value appears in Table G-5, the six-seed geometric-mean bar, and every effect calculation; no seed is removed or imputed. Marker overlap with legend text does not change the values, and Table G-5 supplies exact per-seed readouts.

**Table G-6:** Q9 descriptive absolute summaries aligned with the log-effect estimand

| Mode | S2 geometric mean | S2+Retention geometric mean | Relative effect |
|---|---|---|---|
| ISOLATE | 371.42 | 374.52 | +0.8328% |
| SHARE | 513.70 | 592.07 | +15.2563% |

Throughputs are output tokens/s on the ledger timebase. The effect column uses the ratio of geometric means and equals the back-transformed mean paired log ratio. It is not the ratio of arithmetic mean throughputs.

**Recorded ties.** The ledger throughputs of the ISOLATE pairs at seeds 88002–88004 are equal; the SHARE pair at seed 88005 is also equal. These are recorded ties, not missing values: completed token totals and integer-second ledger durations coincide. The corresponding request-span throughputs may differ slightly. Seed points must not be dropped or interpolated to break such ties.

**Observe-only Q9 audit.** In each S2+Retention–SHARE seed, 48 distinct hashes met producer-side $K = 2$ eligibility and 283–284 unique requests passed the filter; ISOLATE had zero eligible hashes. Across SHARE runs, six RESUME re-entries from four requests had 3,072-token joint-prefix hits. Physical free-skips and adopted-state ownership were not logged, and ISOLATE self-resume was not exercised. Eligibility/hit counts do not prove physical retention. Producer-side $K = 2$ is distinct from preemption-time HPX_RETAIN_SSM/LF_K $= 1$ pinning.

**Scope.** Absolute arm summaries make policy differences easier to read but do not replace the preregistered interaction test. ISOLATE is a compound intervention that can change state creation and memory pressure as well as cross-request reuse. Comparing the SHARE and ISOLATE bar heights directly therefore does not isolate a single reuse mechanism.

Source: supplied `Q9_ABSOLUTE_THROUGHPUT_VALUES.csv` and audit report. The extract contains 48 rows: 24 arm–seed cells on each of the ledger and request-span timebases. This table displays only ledger values; the source CSV retains both timebases.

## G.4 Q9 Effect Estimands, Intervals, and Supplement Navigation

Let $G_{s,p,m} > 0$ denote throughput for seed $s$, policy $p \in \{\mathrm{S2}, \mathrm{Ret}\}$, and mode $m \in \{\mathrm{ISOLATE}, \mathrm{SHARE}\}$. The bar summary and within-mode log effect are

$$\widetilde{G}_{p,m} = \exp\left(\frac{1}{6}\sum_{s} \ln G_{s,p,m}\right), \tag{G1}$$

$$d_m(s) = \ln \frac{G_{s,\mathrm{Ret},m}}{G_{s,\mathrm{S2},m}}, \qquad k(s) = d_{\mathrm{SHARE}}(s) - d_{\mathrm{ISOLATE}}(s). \tag{G2}$$

For $z(s) = d_m(s)$ or $k(s)$, the reported percentage and its frozen log-space paired-$t$ interval are

$$\mathrm{effect}(\%) = 100\big(\exp(\bar{z}) - 1\big), \tag{G3}$$

$$\mathrm{CI}_{95\%}(\%) = 100\left[\exp\left(\bar{z} \pm 2.5705818366147395\,\frac{s_z}{\sqrt{6}}\right) - 1\right]. \tag{G4}$$

Here $s_z$ is the sample standard deviation across the six matched seeds ($df = 5$). For $z = k$, the reported effect is $K_{\mathrm{ret}}$; it is not the arithmetic subtraction of the two displayed percentages. The calculation retains the source campaign's exploratory interval status and does not introduce a new test.

**Table G-7:** Q9 reported effects reproduced from the supplied absolute throughputs

| Timebase | Estimand | Effect (%) | 95% interval (%) | Role |
|---|---|---|---|---|
| ledger | $\Delta_{\mathrm{SHARE}}$ | +15.2563 | $[+0.7993, +31.7867]$ | Secondary |
| ledger | $\Delta_{\mathrm{ISOLATE}}$ | +0.8328 | $[-1.3181, +3.0306]$ | Secondary |
| ledger | $K_{\mathrm{ret}}$ | +14.3043 | $[-0.1057, +30.7930]$ | Primary |
| span | $\Delta_{\mathrm{SHARE}}$ | +18.8179 | $[+2.0241, +38.3760]$ | Sensitivity |
| span | $\Delta_{\mathrm{ISOLATE}}$ | -0.0503 | $[-0.1875, +0.0872]$ | Sensitivity |
| span | $K_{\mathrm{ret}}$ | +18.8776 | $[+1.9768, +38.5794]$ | Sensitivity |

Source frozen method: paired-$t$ in log space, $df = 5$, then percent transformation. These are recomputations from the audit extract, not new observations. "span" denotes request-span sensitivity.

The ledger interaction is positive in all six seeds but its 95% interval includes zero. The retained primary decision is therefore `PERF_CONFIRM_DIRECTIONAL`. The secondary SHARE interval excludes zero; the secondary ISOLATE interval includes zero and does not demonstrate equivalence. The stronger request-span interaction does not replace the ledger primary. Both timebases use the same seed matching, but different time denominators.

**Where the supporting material is located.** Appendix A covers conditional storage models. Appendix B covers environment, measurement, budget rounding, and statistical/adoption scope; B.12 contains the supplied confirmation intervals. Appendix C catalogues campaigns and trace classes. Appendix D.2–D.3 covers implementation invariants, D.4 trace completion, and D.5 numerical and continuation limits. Appendix E contains detailed performance and sensitivity results, with E.6 for sharing and E.14 for run adoption. Appendix F preserves state-history provenance. Appendix G restores the compact result tables and figure-reading qualifications introduced here. These are the same Appendices A–G supplied separately with TPDS and appended to the long version; the former S2–S6 references are not a separate document set.

Reproducibility limit: the delivered Q8/Q9 audit extracts and frozen-method recomputations have been checked, but the original per-request raw logs are not bundled here. Figure 3(b) matched-seed ratios are reconstructed from archived seedwise contrasts, and Q7 precision is reconciled on the 512-token grid in G.1–G.2; neither required a new GPU run.


\end{document}